\documentclass{jfm}
\usepackage{graphicx}
\usepackage{epstopdf, epsfig}
\usepackage{tikz}
\usepackage{tikz-3dplot}
\usepackage{wrapfig}
\usepackage{amsmath, amsfonts}
\usepackage{xcolor}

\def\bk{ {\bf k} }
\def\bq{ {\bf q} }
\def\bp{ {\bf p} }

\newcommand{\Ro}{\mathrm{Ro}}
\def\NOTE#1{\textcolor{black}{#1}}
\def\FINALNOTE#1{\textcolor{black}{#1}}

\shorttitle{Critical transition in fast-rotating turbulence within highly elongated domains}
\shortauthor{A. van Kan and A. Alexakis}

\title{Critical transition in fast-rotating turbulence within highly elongated domains }

\author{Adrian van Kan\aff{1}
  \corresp{\email{adrian.van.kan@phys.ens.fr}},
  \and  Alexandros Alexakis \aff{1}}

\affiliation{\aff{1} Laboratoire de Physique de l'Ecole normale sup\'erieure, ENS,
Universit\'e PSL, CNRS, Sorbonne Universit\'e, Universit\'e Paris-Diderot,
Sorbonne Paris Cit\'e, Paris, France}
\begin{document}

\maketitle

\begin{abstract}
\NOTE{We study rapidly rotating turbulent flows in a highly elongated domain 
using an asymptotic expansion at simultaneously low Rossby number $Ro\ll 1$ and large domain height compared to the energy injection scale, $h=H/\ell_{in}\gg 1$.
We solve the resulting equations using an extensive set of 
direct numerical simulations for different parameter regimes.
As the parameter $\lambda = (h Ro)^{-1}$ is increased beyond a threshold $\lambda_c$, a transition is observed from a state without an inverse energy cascade to a state with an inverse energy cascade.  }
For large Reynolds number and large horizontal box size, we provide evidence for criticality of the transition in terms of the large-scale energy dissipation rate.
\end{abstract}


\section{Introduction}
\label{sec:intro}
Rotating fluid flows are commonly encountered in astrophysical and geophysical systems such as planetary and stellar interiors, planetary atmospheres and oceans \citep{pedlosky2013geophysical}, as well as in industrial processes involving rotating machinery. The fluid motions in these systems are typically turbulent, i.e. the Reynolds number $\Rey$, which is defined as the ratio between inertial and viscous forces, is large. At the same time the flow is affected by the Coriolis force due to system rotation. The magnitude of the Coriolis force compared to the inertial force is measured by the nondimensional Rossby number $\Ro=U /(\Omega \ell)$, where $\Omega$ is the rotation rate and $U$ and $\ell$ are typical velocity and length scales of the flow. For $\Ro<\infty$, the isotropy of classical three-dimensional (3-D) turbulence is broken, since the rotation axis imposes a direction in space. When the rotation rate is large, i.e. in the limit $\Ro \to 0$, the rotation tends to suppress variations of the motion along the axis of rotation and thus makes the flow quasi-two-dimensional, an effect described by the Taylor-Proudman theorem \citep{ hough1897ix, proudman1916motion, taylor1917motion,greenspan1968theory}. 

As is well known, the properties of turbulent cascades strongly depend on the dimension of space. In homogeneous isotropic 3-D turbulence, energy injected at large scales is transferred, by non-linear interactions, to small scales in a \textit{direct energy cascade} \citep{frisch1995turbulence}. In the two-dimensional (2-D) Navier-Stokes equations both energy and enstrophy are inviscid invariants and this fact constrains the energy transfer to be from small to large scales in an \textit{inverse energy cascade} \citep{bofetta2012twodimensionalturbulence}. 
When $\Ro$ is lowered below a certain threshold value $\Ro_c$ in a rotating turbulent flow, a transition is encountered where the flow becomes quasi-2-D and an inverse cascade develops.
In this state, part
of the injected energy cascades to larger scales and another part to smaller scales, forming what is referred to as a {\it split} or {\it bidirectional} cascade, see \citep{alexakis2018cascades}.
In the absence of an effective large-scale damping, this inverse cascade can lead to the formation of a condensate in which the energy is concentrated at the largest scale.

The formation of large-scale quasi-2-D structures in rotating flows
has been observed early on in experiments \citep{ibbetson1975experiments, hopfinger1982turbulence, dickinson1983oscillating} and numerical simulations \citep{bartello1994coherent, yeung1998numerical, godeferd1999direct, smith1999transfer}. Since then, various investigations have focused on different aspects of the quasi-2-D behaviour of rotating turbulence experimentally \citep{baroud2002anomalous, baroud2003scaling, morize2006energy, staplehurst2008structure, van2009experiments, duran2013turbulence, yarom2013experimental, machicoane2016two} and numerically \citep{ mininni2009scale, thiele2009structure, favier2010space,mininni2010rotating, sen2012anisotropy, marino2013inverse, biferale2016coherent, valente2017spectral, buzzicotti2018energy, buzzicotti2018inverse}.
In particular, recent experiments were able to investigate the presence of the inverse cascade \citep{yarom2014experimental, campagne2014direct, campagne2015disentangling, campagne2016turbulent}. The transition from a forward to an inverse cascade in rotating turbulence was studied systematically using numerical simulations in \citep{smith1996crossover,deusebio2014dimensional,pestana2019regime,pestana2020rossby}, while the transition to a condensate regime was studied in \citep{alexakis2015rotating, yokoyama2017hysteretic, seshasayanan2018condensates}.

Similar transitions from a forward to an inverse cascade and to quasi-2-D motion have also been observed in other systems like thin-layer turbulence \citep{celani2010morethantwo, benavides_alexakis_2017, musacchio2017split,  vankan2019condensates,  musacchio2019condensate, vankan2019rare}, stratified turbulence \citep{sozza2015dimensional}, rotating and stratified flows \citep{marino2015resolving}, magneto-hydrodynamic systems \citep{alexakis2011two, seshasayanan2014edge, seshasayanan2016critical} and helically constrained flows \citep{sahoo2015disentangling, sahoo2017discontinuous} among others (see the articles by \cite{alexakis2018cascades} and \cite{pouquet2019helicity} for recent reviews).

While the existence of a transition from forward to inverse energy cascade is well-established in many systems, including rotating turbulence, its detailed properties remain poorly understood in most cases. Turbulent flows involve non-vanishing energy fluxes and thus are out-of-equilibrium phenomena \citep{Goldenfeld2017turbnonequil}. 
While in the case of the laminar-turbulence transition in shear flows, a connection with non-equilibrium statistical physics has been established by placing the problem in the directed percolation universality class \citep{pomeau1986front, manneville2009spatiotemporal}, in particular for plane Couette flow \citep{lemoult2016directed,chantry2017universal} and pipe flow \citep{moxey2010distinct}, such a general theoretical link remains yet to be found for the non-equilibrium transition from forward to inverse energy cascade. However, previous numerical studies have successfully analysed special cases. For instance in the case of thin-layer turbulence, \cite{benavides_alexakis_2017} were able to provide strong evidence for criticality of the inverse energy transfer rate as a function of a control parameter related to box height at the transition to an inverse cascade. The term criticality is used here to describe situations where an order parameter (e.g. the rate of inverse energy transfer) changes from zero to non-zero at a critical value of a control parameter (e.g. box height, $\Ro$). When the limit of infinite horizontal box size and $\Rey\to \infty$ is taken this change can be either discontinuous (1st order) or continuous with discontinuous (first/second/higher) derivative (2nd order) at the critical point, (for a more detailed discussion, see \citep{alexakis2018cascades}). Knowing whether the transition to an inverse cascade in a turbulent flow is critical or smooth is important, in particular since this information is paramount for further investigations. 
For instance, in a critical transition, two separated phases exist and one may meaningfully speak of the phase diagram of the system. This is particularly interesting in situations with many parameters, such as rotating stratified turbulence in finite domains. Furthermore, near the critical points there are critical exponents to be measured, for which a comparison with theoretical predictions seems possible.

In the case of rotating turbulence in a layer of thickness $H$ (after the limit of infinite horizontal box size $L$ and $\Rey$ is taken) there are two control parameters left as a function of which the system can display criticality:
the ratio $h=H/\ell_{in}$ (where here $\ell_{in}$ is taken to be the forcing length scale)
and $Ro$. If criticality is present, then this 2-D space ($h,Ro$) will be split into two regions, in one of which an inverse cascade is observed, but no inverse cascade in the other. The two regions are separated by a critical line given by $h_c(Ro)$ that needs to be determined. For large $Ro$ (weak rotation), the problem reduces to that of the non-rotating layer and therefore $\lim_{Ro\to \infty} h_c(Ro) = h_c^* > 0$, where $h_c^*$ is the critical value of $h$ for the non-rotating layer \citep{celani2010morethantwo, benavides_alexakis_2017, vankan2019condensates}. For small $Ro$, the scaling of $h_c$ with $Ro$ is not known. \cite{deusebio2014dimensional} investigated this problem and showed evidence for a continuous transition, with $h_c$ increasing as $Ro$ was decreased, but could not reach small enough $Ro$ to determine a scaling of $h_c$ with $Ro$. 
In \citep{alexakis2018cascades} it was argued that the scaling $h_c \propto 1/Ro$ should be 
followed, but so far no evidence numerical or experimental exists to support or dismiss this conjecture.
This is what we address 
in this work by studying the simultaneous limit of asymptotically small $Ro$ and large domain height.

The remainder of this paper is structured as follows. In section \ref{sec:theo_bg} we discuss the theoretical background of this study, in  section \ref{sec:setup}, we introduce the set-up of our numerical simulations and define the quantities to be measured. In section \ref{sec:results}, we describe the results of the direct numerical solutions (DNS) we performed and finally in section \ref{sec:conclusions}, we draw our conclusions and discuss remaining open problems.

\section{Theoretical Background}
\label{sec:theo_bg}

\subsection{Quasi-two-dimensionalisation and \NOTE{inertial waves}}
\label{subsec:quasi2D_wt}
In this section we discuss the theoretical results underpinning the present study. A fundamental property of rotating flows is the fact that they support inertial wave motions, whose restoring force is the Coriolis force \citep{greenspan1968theory}. Inertial waves have the peculiar anisotropic dispersion relation
\begin{equation}
\omega^{s_\bk}(\mathbf{k}) = 2s_{\bf k} \Omega k_\parallel/k, \label{eq:iw_dr}
\end{equation}
where $s_\bk=\pm 1$, $\Omega$ is the rotation rate, $k_\parallel$ is the component of $\mathbf{k}$ along the rotation axis and $k=|\mathbf{k}|$. Similarly, we define $k_\perp$ as the magnitude of the component of $\mathbf{k}$ perpendicular to the rotation axis. In the remainder of this article, \textit{parallel} and \textit{perpendicular} will always refer to the rotation axis. Inertial waves in fast-rotating turbulence are important for understanding the direction of the energy cascade, as will be discussed below. The form of (\ref{eq:iw_dr}) shows that motions which are invariant along the axis of rotation, i.e. which are 2-D with three components (2D3C), have zero frequency and are thus unaffected by rotation. This allows decomposing the flow into two components, the 2D3C modes which are not directly affected by rotation, forming the \textit{slow manifold}, and the remaining 3-D modes which are affected by the rotation, forming the \textit{fast manifold} \citep{buzzicotti2018inverse}. In the limit $\Ro \to 0$, it can be shown that only resonant interactions remain present \citep{waleffe1993inertial,embid1998low,chen2005resonant}. Resonant interactions are those interactions between wavenumber triads $(\mathbf{k},\mathbf{p},\mathbf{q})$ satisfying 
\begin{eqnarray}
  \mathbf{k} + \mathbf{p} +\mathbf{q} =& 0 \label{eq:res_kpq}, \\\
  \omega^{s_\bk}( \mathbf{k}) +\omega^{s_\bp}( \mathbf{p}) + \omega^{s_\bq}(\mathbf{q}) =& 0, \label{eq:res_om}
\end{eqnarray}
where $\omega^{s_\bk}(\bk), \omega^{s_\bp}(\bp)$ and $\omega^{s_\bq}(\bq)$ are given by (\ref{eq:iw_dr}).
When only resonant interactions are present in the system, it can further be shown that any triad including modes from both the fast and slow manifolds leads to zero net energy exchange between the two manifolds. Thus, with only resonant interactions, the slow and fast manifolds evolve independently from each other without exchanging energy, and there is inverse energy transfer in the perpendicular components of the slow manifold. This decoupling may lead to an inverse energy cascade for the quasi-2-D part of the flow. 
In fact, it can be proven that for finite Reynolds number $Re\equiv U\ell /\nu$ (where $\nu$ is viscosity, $U$ is r.m.s. velocity and $\ell$ is a forcing length scale) and finite $H$, the flow will become exactly 2-D
as $Ro\to 0$ \citep{gallet2015exact}. 

On the other hand, in the limit of large domain height $H$, very small values of $k_\parallel$ are possible, such that quasi-resonant triads, for which (\ref{eq:res_om}) is only satisfied to $O(\Ro)$, can transfer energy between the slow and fast manifolds. Thus the inverse energy transfer in the slow manifold may be suppressed by interaction with quasi-resonant 3-D modes. Asymptotically, for infinite domains and $k_\parallel / k_\perp \ll 1$, wave turbulence theory predicts a forward energy cascade and an associated anisotropic energy spectrum \citep{galtier2003weak}. 

There are thus two mechanisms at play in the energy transport: the dynamics of the slow manifold transferring energy to the large scales and the 3-D interactions transferring energy to the small scales. 
Which of these two processes dominates depends on the two nondimensional parameters, the ratio $h=H/\ell_{in}$, where $\ell_{in}$ is the forcing length scale, and the Rossby number $\Ro=\epsilon_{in}^{1/3} / (\ell_{in}^{2/3} \Omega)$ based on $\ell_{in}$ and the forcing velocity scale $(\epsilon_{in}\ell_{in})^{1/3}$ resulting from the energy injection rate $\epsilon_{in}$. 
The main criterion is whether or not 2-D modes are isolated from 3-D modes due to fast rotation. 
The coupling of 2-D and 3-D motions will be strong enough to stop the inverse cascade if the fast modes closest to the slow manifold ($k_\parallel \sim  H^{-1}$, $k_\perp \sim \ell_{in}^{-1}$) 
are `slow' enough to interact with the 2-D slow manifold. This implies that
their wave frequency $\omega = 2\Omega k_\parallel/k_\perp \sim 2\Omega \ell_{in}/H$
is of the same order as the non-linear inverse time scale $\tau_{nl}^{-1} \sim \epsilon_{in}^{1/3} \ell_{in}^{-2/3}$. This leads to the following prediction for the critical height $H_c$, where the transition takes place,
\begin{equation}
h_c=\frac{H_c}{\ell_{in}} \propto \Omega \epsilon_{in}^{-1/3} \ell_{in}^{2/3} = \Ro^{-1}.  \label{eq:transition_forward_inverse}   
\end{equation}
Importantly, the predicted critical rotation rate and height are linearly proportional. The criterion (\ref{eq:transition_forward_inverse}) suggests that the nondimensional control parameter of the transition in the limit of large $h$ and small $\Ro$ is given by 
\begin{equation}
    \lambda = \frac{1}{h \times \Ro} = \frac{\ell_{in}^{5/3} \Omega }{\epsilon_{in}^{1/3} H}.
    \label{eq:def_lamb}
\end{equation}

\NOTE{ 
\subsection{Multiscale expansion}
\label{subsec:multiscale} }
In the present paper, we will explore the regime of simultaneously large $h$ and small $\Ro$. Brute-force simulations at small $\Ro$ are costly, since very small time steps are required to resolve the fast waves of interest. Rather, we exploit an asymptotic expansion first introduced in \citep{julien1998new}, which allows to test the prediction (\ref{eq:transition_forward_inverse}) and to investigate the properties of the transition to a split cascade. The expansion is based on the constant-density Navier-Stokes equation in a system rotating at the constant rate $\mathrm{\boldsymbol{\Omega}} = \Omega \hat{e}_\parallel$, which in its dimensional form reads
\NOTE{
\begin{eqnarray}
    \partial_t \mathbf{u} + \mathbf{u}\cdot \nabla \mathbf{u} + 2\Omega \hat{e}_\parallel\times\mathbf{u}=& - \nabla p + \nu \nabla^2\mathbf{u}+ \mathbf{f} \label{eq:RNS_mt}, \\
    \nabla \cdot \mathbf{u} =&0 \label{eq:inc_mt},
\end{eqnarray} }
with time $t$, velocity $\mathbf{u}$, pressure (divided by constant density) $p$, kinematic viscosity $\nu$, and the forcing $\mathbf{f}$. The domain considered here is the cuboid of dimensions 
\begin{wrapfigure}{l}{0.305\textwidth}
 \centering
  \includegraphics[height=5.5cm]{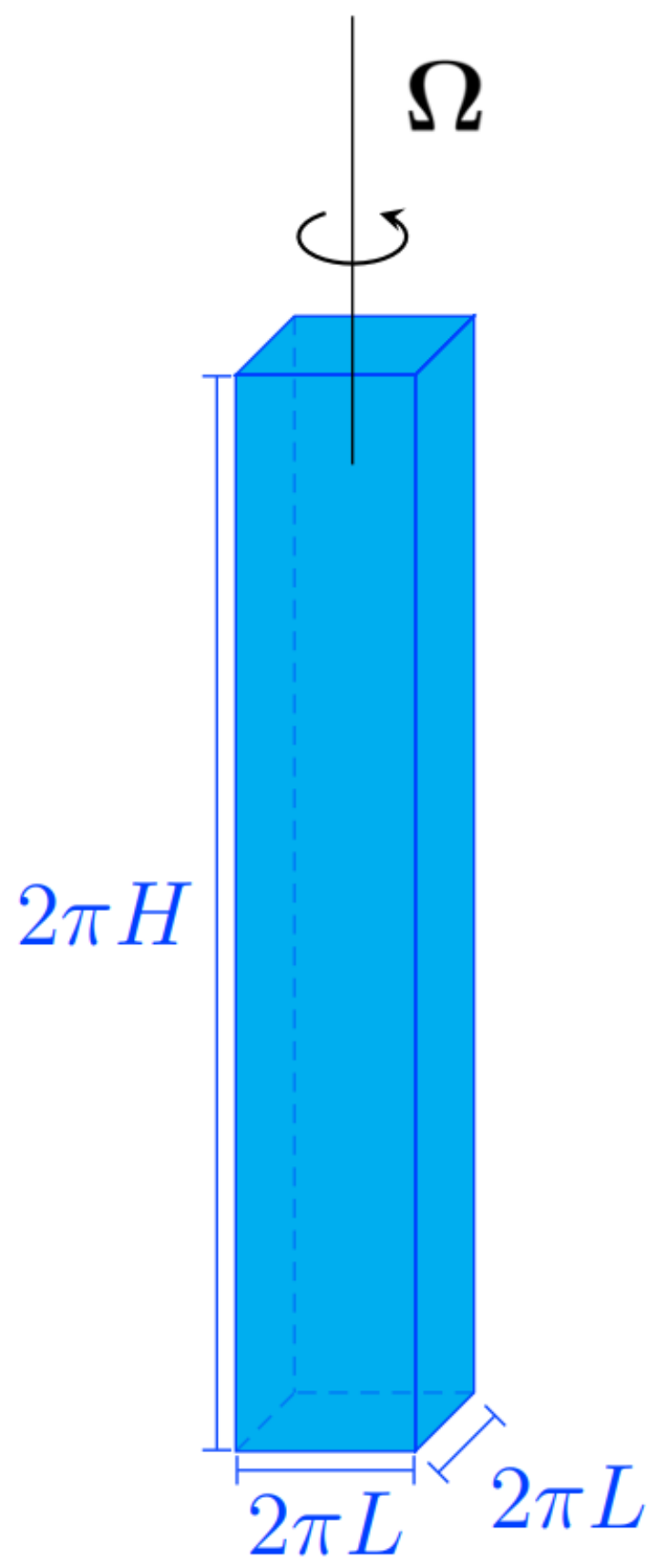}
  \caption{The long, rapidly rotating box domain.} \label{fig:LRRB}
\end{wrapfigure}
\noindent $2\pi L\times2 \pi L \times2 \pi H$, depicted in figure \ref{fig:LRRB}, with periodic boundary conditions. For any vector $\mathbf{F}$, we define the parallel and perpendicular components as $\mathbf{F}_\parallel = (\mathbf{F}\cdot \hat{e}_\parallel) \hat{e}_\parallel = F_\parallel \hat{e}_\parallel$ and $\mathbf{F}_\perp = \mathbf{F}-\mathbf{F}_\parallel$. 

We choose to consider a stochastic forcing injecting energy at a constant mean rate into both perpendicular and parallel motions $\langle \mathbf{f}_\perp \cdot \mathbf{u}_\perp \rangle =\langle f_\parallel u_\parallel  \rangle = \epsilon_{in}/2 \Rightarrow \langle \mathbf{f}\cdot\mathbf{u}\rangle  = \epsilon_{in}$, where $\langle \cdot \rangle$ denotes an ensemble average over inifinitely many realisations. The forcing is chosen to be 2-D (independent of the parallel direction), for simplicity, and filtered in Fourier space to act only on a ring of perpendicular wavenumbers $\mathbf{k}$ centered on $|\mathbf{k}| = k_f=1/\ell_{in}$. A similar 2-D forcing has been widely used in previous studies on the transition toward an inverse cascade, such as \citep{smith1996crossover, celani2010morethantwo, deusebio2014dimensional}. 
In the present case, it makes sense restricting the forcing to the 2-D modes because for any 
forcing with finite correlation time, the injection of energy to the $k_\|\ne0$ modes would be suppressed when the limit $\Omega\to\infty$ is taken due to the high wave frequencies. Thus, most of the energy would be injected into the $k_\|=0$ modes. We note, however, that in general the transition to an inverse cascade can depend on the choice of forcing. 
Recent work in thin-layer turbulence by \cite{poujol2020role} suggests that a 3-D forcing, which includes non-zero parallel wavenumbers, is less efficient at generating an inverse cascade and delays the onset. \FINALNOTE{A related problem of interest, which we do not address in the present study, is to investigate the transfer of energy to the 2-D manifold in the case when only the 3-D modes are forced. This has recently been studied experimentally and theoretically \cite{le2019experimental, brunet2020shortcut,reun2020near}.  
}


The forcing imposes a time scale $(\ell_{in}^2/\epsilon_{in})^{1/3}$ as well as a length scale $\ell_{in}$, and thus a velocity scale $(\epsilon_{in} \ell_{in})^{1/3}$. However, the typical scale of parallel variations is $H$, rather than $\ell_{in}$. As detailed in appendix \ref{sec:appA}, nondimensionalisation using these scales reveals that the nondimensional control parameters are indeed the Rossby number $Ro = (\epsilon_{in} \ell_{in})^{1/3}/(\Omega \ell_{in})$ and the rescaled domain height $h=H/\ell_{in}$, in addition to the Reynolds number $\Rey = (\epsilon_{in}\ell_{in}^4)^{1/3}/\nu$ and the rescaled domain width $\Lambda = L/\ell_{in}$. We consider tall boxes $h = \epsilon^{-1}\gg 1$, under the influence of fast rotation, $Ro=O(\epsilon)\ll 1$, such that $\lambda =(h Ro)^{-1}= O(1)$ (independent of $\epsilon$), while $\Rey=O(1)$ and $\Lambda=O(1)$. In particular, the expansion assumes that $\Rey \ll Ro^{-1}$, $ \Rey \ll H/\ell_{in}$, $L/\ell_{in} \ll Ro^{-1}$ and $L \ll H$. 
\NOTE{We note that this limiting procedure does not correspond to the weak turbulence limit, for which the limit $h\to \infty$
is taken before the limit $\Ro\to0$.}
The method of multiple scales or a heuristic derivation (see appendix \ref{sec:appA}) can be used to obtain a set of asymptotically reduced equations for the parallel components of velocity $u_\parallel$ and vorticity $\omega_\parallel = (\nabla\times \mathbf{ u})\cdot \hat{e}_\parallel$ whose dimensionless form reads
\begin{eqnarray}
    \partial_t u_\parallel + \mathbf{u}_\perp\cdot\nabla_\perp u_\parallel &+ 2\lambda \partial_\parallel\nabla^{-2}_\perp\omega_\parallel &= \frac{1}{\Rey} \nabla^2_\perp u_\parallel+ f_\parallel, \label{eq:red_vert_mom} \\
    \partial_t \omega_\parallel +\mathbf{u}_\perp \cdot \nabla_\perp \omega_\parallel &- 2\lambda \partial_\parallel  u_\parallel &= \frac{1}{\Rey} \nabla_\perp^2\omega_\parallel+ f_\omega, \label{eq:red_vert_vort}
\end{eqnarray}
where $\partial_\parallel$ is the partial derivative in the parallel direction, $\nabla_\perp = \nabla-\hat{e}_\parallel \partial_\parallel$ and $f_\omega = (\nabla\times \mathbf{f})\cdot\hat{e}_\parallel$. The perpendicular components $\bf u_\perp$ are divergence-free 
to leading order, $\bf \nabla_\perp\cdot u_\perp =0 $, which permits us to
write them in terms of a stream function, ${\bf u}_\perp = \hat{e}_\parallel \times \nabla \psi$, where $\psi$ is such that $\omega_\parallel = \nabla_\perp^2 \psi $. These nondimensional asymptotic equations are valid in the domain $2\pi \Lambda \times 2 \pi \Lambda \times 2 \pi$. Importantly, in equations (\ref{eq:red_vert_mom}) and (\ref{eq:red_vert_vort}), all the information about $H$ and $\Omega$ is contained in the parameter $\lambda$, which is defined by (\ref{eq:def_lamb}). This implies that if a transition from a direct to a split energy cascade is captured in these asymptotic equations, the single control parameter of the transition indeed is given by $\lambda$ (in the limit of large $Re$ and $\Lambda$), as predicted in by the heuristic arguments in section \ref{subsec:quasi2D_wt}.
Variants of the asymptotic equations (\ref{eq:red_vert_mom}, \ref{eq:red_vert_vort}) have been extensively used in the past, in particular for studying rotating turbulence \citep{nazarenko2011critical} and rapidly rotating convection (adding energy equation) \citep{sprague2006numerical,julien2012statistical,julien2012heat,rubio2014upscale, grooms2010model}, as well as dynamos driven by rapidly rotating convection (adding the energy and MHD induction equations) \citep{calkins2015multiscale}.

The equations (\ref{eq:red_vert_mom}) and (\ref{eq:red_vert_vort}) are closely related to well-known models in geophysical fluid dynamics. In particular, since the leading-order perpendicular velocity is in geostrophic balance and advection is purely perpendicular, the model bears a resemblance to the classical quasi-geostrophic (QG) approximation valid in thin layers, see e.g. \cite{pedlosky2013geophysical}. Indeed, equations (\ref{eq:red_vert_mom}, \ref{eq:red_vert_vort}) have been referred to as \textit{generalised QG equations} \citep{julien2006generalized}. A great advantage of the reduced equations over the full Navier-Stokes equations is that they can be efficiently integrated numerically, as explained below. Note that in the expansion, as a consequence of the Taylor-Proudman constraint applied in the limit $\Ro\to 0$, $h\to \infty$, fast variations in the parallel direction are eliminated, i.e. $k_\parallel \ll k_\perp$ in terms of dimensional wavenumbers. \NOTE{Equations (\ref{eq:red_vert_mom}) and (\ref{eq:red_vert_vort}) retain inertial waves with the dispersion relation, in nondimensional form,
\begin{equation}
\omega(\mathbf{k}) = \pm 2 \lambda \frac{k_\parallel}{k_\perp}. \label{eq:disprel_red}
\end{equation}
The rescaled wavenumbers $k_\perp \geq 1/\Lambda$ and $k_\parallel\geq 0$ and $\lambda$ are all $O(1)$ (independent of $\epsilon$), therefore only the inertial waves with order one frequencies, i.e. those on the parallel scale of the layer depth, are retained in the reduced equations. Fast inertial waves, i.e. those whose parallel scale is comparable to the perpendicular scale and whose frequencies are comparable to the rotation rate $\Omega$, are filtered out.} For this reason, the asymptotic reduction gives a significant improvement in efficiency (the filtering of fast inertial waves here is similar to the filtering of fast inertia-gravity waves in classical QG). We perform direct numerical simulations (DNS) of (\ref{eq:red_vert_mom}) and (\ref{eq:red_vert_vort}) to show that, as predicted by the theory outlined above, there is indeed a transition from a direct to an inverse energy cascade in this extreme parameter regime.

\NOTE{
\subsection{A homochiral triadic instability}  
\label{subsec:homoch_instab}  }                 
In this subsection, we discuss a linear instability mechanism present in the asymptotically reduced governing equations (\ref{eq:red_vert_mom},  \ref{eq:red_vert_vort}), which will be helpful for the interpretation of the DNS results. For concreteness, we work in the canonical basis and choose $\hat{e}_\parallel=\hat{e}_z$. The Fourier transformed governing equations then read, in the absence of forcing or dissipation,
\begin{eqnarray}
    \partial_t\hat{u}_\parallel(\mathbf{k}) - 2 i \lambda\frac{k_\parallel}{k_\perp^2} \hat{\omega}(\mathbf{k}) =&- \sum_{\mathbf{p}+\mathbf{q}+\mathbf{k}=0} [p_xq_y-q_xp_y] \frac{\hat{\omega}_\parallel^*(\mathbf{p})}{p_\perp^2}  \hat{u}_\parallel^*(\mathbf{q}), \label{eq:vertvel_Fourier} \\
    \partial_t\hat{\omega}_k - 2i \lambda k_\parallel \hat{u}_k =& -\sum_{\mathbf{p}+\mathbf{q}+\mathbf{k}=0} [p_xq_y-q_xp_y]\frac{\hat{\omega}_\parallel^*(\mathbf{p})}{p_\perp^2} \hat{\omega}_\parallel^*(\mathbf{k}) , \label{eq:vertvort_Fourier}
\end{eqnarray}
where it was used that $u_\parallel \equiv\sum_\mathbf{k}\hat{u}_\parallel(\mathbf{k}) e^{i \mathbf{k}\cdot \mathbf{x}}$, similarly $\omega_\parallel \equiv \sum_\mathbf{k}\hat{u}_\parallel(\mathbf{k}) e^{i \mathbf{k}\cdot \mathbf{x}}$, in addition to $ \omega_\parallel = \nabla_\perp^2 \psi$, and $\mathbf{u}_\perp = \hat{e}_\parallel \times \nabla \psi$. \NOTE{It is helpful to reformulate the dynamics in a helical basis as in \cite{waleffe1993inertial}. The linearisation of (\ref{eq:vertvel_Fourier}, \ref{eq:vertvort_Fourier}) may be written as
\begin{equation}
    \partial_t \mathbf{X}(\mathbf{k}) = L(\mathbf{k}) \mathbf{X}(\mathbf{k}),\hspace{1cm} L(\mathbf{k})= i \omega(\mathbf{k}) \begin{pmatrix} 0 &1 \\ 1 & 0  \end{pmatrix}
\end{equation}
with $\mathbf{X}(\mathbf{k})=\left(\hat{u}_\parallel(\mathbf{k}), \omega_\parallel(\mathbf{k})/k_\perp\right)$ and $\omega(\mathbf{k}) = 2 \lambda k_\parallel/k_\perp$, identical to (\ref{eq:disprel_red}) up to the sign. 
The eigenvectors of $L(\mathbf{k})$ corresponding to the eigenvalues $\pm \omega(\mathbf{k})$ are given by $\left\lbrace \begin{pmatrix} 1 \\ 1\end{pmatrix}, \begin{pmatrix} 1 \\ -1\end{pmatrix} \right\rbrace$. 
%
They correspond to inertial waves of positive and negative helicity respectively, with dispersion given by (\ref{eq:disprel_red}). Representing $\mathbf{X}(\mathbf{k})$ in this eigenbasis leads to the new variables $Z^{s_\mathbf{k}}_\mathbf{k} = \hat{u}_\mathbf{k} + s_\mathbf{k} \frac{\hat{\omega}_\mathbf{k}}{k_\perp}$ where $s_\mathbf{k}=\pm 1$. The full nonlinear system may then be written entirely in terms of the $Z_\mathbf{k}^{s_\mathbf{k}}$.}

We consider a (not necessarily resonant) triad  ($\mathbf{k}$,$\mathbf{p}$,$\mathbf{q}$) of rescaled wavenumbers with $\mathbf{k}=(k_f=1,0,0)$ being the forcing wavenumber, implying $p_x=-k_f-q_x$, $p_y=-q_y$, $p_z=-q_z$. We choose the forcing-scale mode $Z^{+}_\mathbf{k}=u_0$, $Z^{-}_\mathbf{k}=0 \Leftrightarrow \hat{\mathbf{u}}_k =u_0 (0,-i/2,1/2)$, i.e. the positively helical flow $\mathbf{u}=u_0 (0,\sin(k_f x), \cos(k_f x))$. We take the modes at $\mathbf{p}$ and $\mathbf{q}$ to be small-amplitude inertial waves, and perform a linear stability analysis of this configuration for the homochiral case $s_\mathbf{p}=s_\mathbf{q}=1=s_\mathbf{k}$ (the other cases do not give relevant results). \NOTE{We thus determine the growth rate $\sigma(\mathbf{q})$ ($\mathbf{p}$ is uniquely determined by $\mathbf{q}$) of the two inertial-wave modes, whose temporal evolution is given by $Z_\mathbf{q}^+,Z_\mathbf{p}^+\propto \exp(\sigma t)$. }

\begin{figure}
    \centering
    \includegraphics[width=5.0cm]{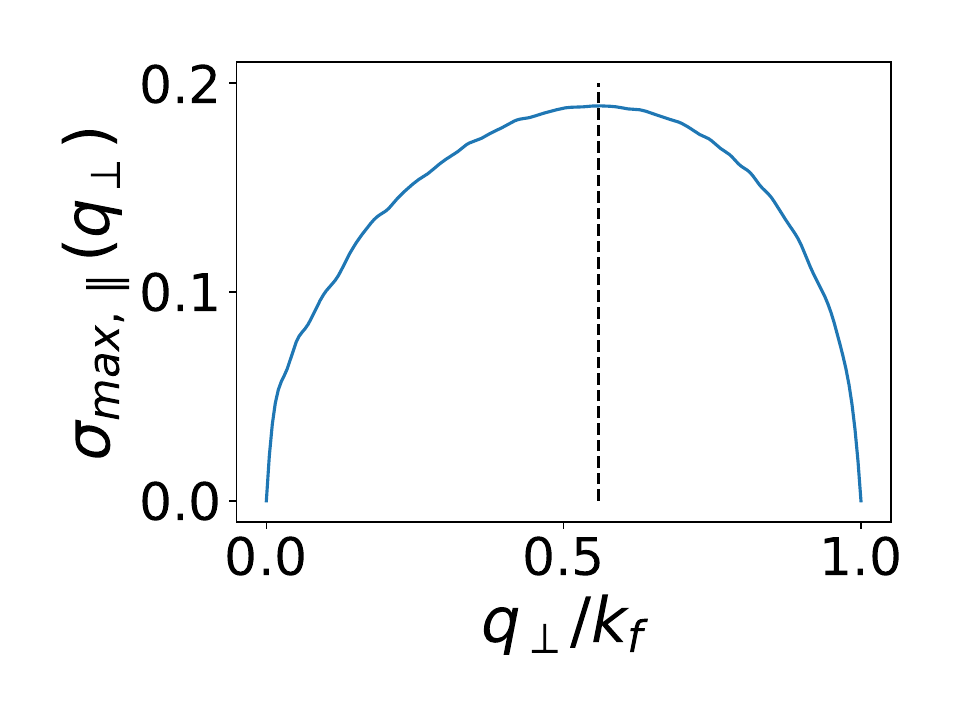}\includegraphics[width=5.0cm]{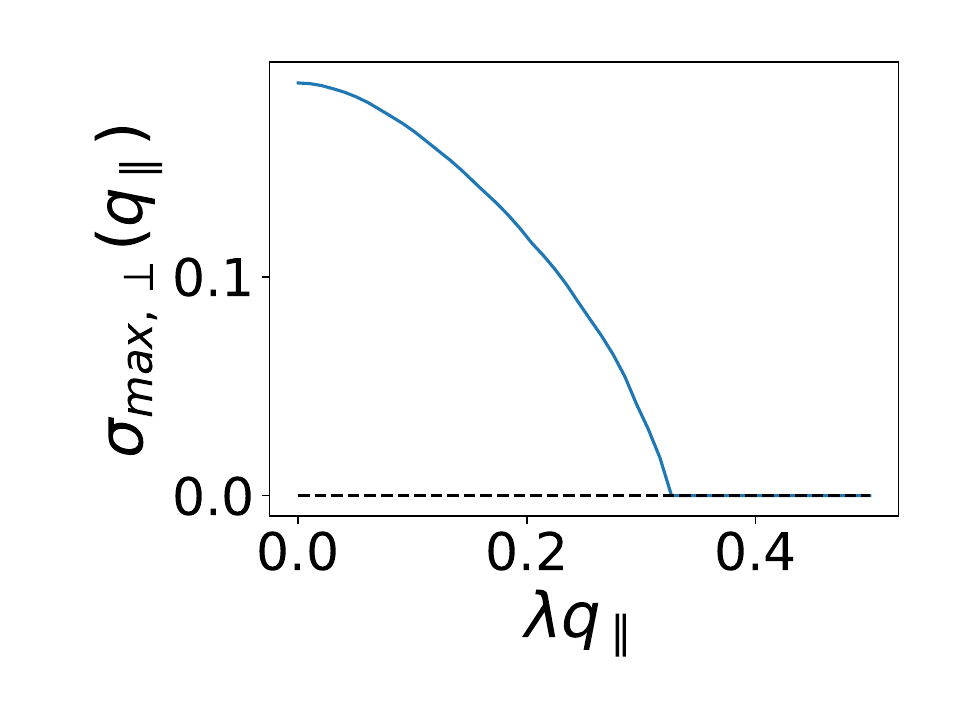}
    \caption{Left: Maximum (over $q_\parallel$) of the real part of the growth rate as a function of $q_\perp$. The maximum is found for $q_\perp \approx k_f/2$. Right: Maximum (over $q_\perp$) of the real part of the growth rate as a function of $\lambda q_\parallel$ with $k_f=1$. There is a monotonic decay with $q_\parallel$ up to $\lambda q_\parallel\approx 0.35$ and vanishing values beyond this point.}
    \label{fig:growth_rate_triad}
\end{figure}

 The left panel of figure \ref{fig:growth_rate_triad} shows that the maximum of $\sigma$ occurs for small wavenumbers at $q_\perp\approx k_f/2$, while the right panel indicates that this maximum is located at $q_\parallel=0$. 
Thus the 2-D base flow becomes unstable to smaller perpendicular wavenumbers $q_\perp$ and for a range of parallel wavenumbers $|q_\parallel| \lesssim 0.35/\lambda $.
Since the rescaled layer height is given by $2\pi$, the minimum parallel wavenumber is $q_\parallel^{min} = 1$. 
Thus the $q_\parallel\ne 0$ modes are unstable only if $\lambda\lesssim 0.35$. For $\lambda$ larger than this value all $q_\parallel\ne 0$ wavenumbers are stable. For large values of $\lambda$ therefore the 2-D modes $q_\parallel=0$ are expected to decouple from the 3-D modes $q_\parallel\ne0$ and an inverse cascade is expected. Conversely, for small values of $\lambda$ the 3-D modes become unstable and can possibly redirect energy back to small scales. We find the signature of this triadic instability in the DNS results which we discuss in section \ref{sec:results}.

\section{Numerical set-up and methodology}  
\label{sec:setup}                           
\NOTE{In this section, we describe the numerical set-up used in the present study. The PDEs which we solve numerically in a domain $2\pi \Lambda\times 2\pi \Lambda \times 2\pi$ are given by (\ref{eq:red_vert_mom}) and (\ref{eq:red_vert_vort}) with modified dissipative terms,}
\begin{eqnarray}
    D_t^\perp  u_\parallel + & 2\lambda \partial_\parallel\nabla^{-2}_\perp\omega_\perp &= - \frac{(-\nabla^{2}_\perp)^n u_\parallel}{\Rey_\nu}-\frac{(-\nabla_\parallel^{2})^m u_\parallel}{\Rey_\mu} - \frac{u_\parallel^{ls}}{\Rey_\alpha} +  f_\parallel, \label{eq:red_vert_mom_dissmod} \\
    D_t^\perp \omega_\parallel - &2\lambda \partial_\parallel u_\parallel &= - \frac{(-\nabla^{2}_\perp)^n \omega_\parallel}{\Rey_\nu}-\frac{(-\nabla_\parallel^{2})^m \omega_\parallel}{\Rey_\mu} - \frac{\omega_\parallel^{ls}}{\Rey_\alpha} +f_\omega. \label{eq:red_vert_vort_dissmod}
\end{eqnarray}
Here $D_t^\perp = \partial_t + \mathbf{u}_\perp \cdot \nabla_\perp$. 
The right-hand side of eqs. (\ref{eq:red_vert_mom_dissmod},\ref{eq:red_vert_vort_dissmod})
expresses the dissipation terms and the forcing. For a field $g$ we define $g^{ls}=\sum_{\mathbf{k}, k_\perp \leq 2} \hat{g}(\mathbf{k}) \exp(i\mathbf{k}\cdot \mathbf{x})$, in terms of the Fourier transform $\hat{g}(\mathbf{k})$ of $g$ with $\mathbf{k}\in \mathbb{N}^3$. The large-scale friction terms involving $u_\parallel^{ls}$ and $\omega_\parallel^{ls}$ have been added to prevent the formation of a condensate at small wavenumbers. A technical advantage of this type of large-scale friction over more commonly used hypodissipation is that it only directly affects the small wavenumbers, which are to be damped. \NOTE{The term proportional to $\nabla_\parallel^{2m}(\cdot)$ has been artificially added to the equations, it does not appear in the asymptotically reduced equations (\ref{eq:red_vert_mom}), (\ref{eq:red_vert_vort}), since it is asymptotically small. It has nonetheless been added to suppress exceedingly large parallel wave-numbers which are expected not to interact significantly with the slow manifold, thereby reducing the required resolution in the parallel direction and the computational cost. }
The hyperviscosity exponents $n=4$ and $m=2$ were used in all simulations. 

The resulting equations (\ref{eq:red_vert_mom_dissmod}, \ref{eq:red_vert_vort_dissmod}) contain five nondimensional parameters. First, $\Lambda$ stemming from the boundary conditions and $\lambda$ defined in equation (\ref{eq:def_lamb}). In addition, there are three different Reynolds numbers based on the three dissipation mechanisms $\Rey_\nu = \epsilon_{in}^{1/3} \ell^{2n-2/3}/\nu_n$, $\Rey_\mu=\epsilon_{in}^{1/3} \ell^{2m-2/3}/\mu_m$ and $\Rey_\alpha=\epsilon_{in}^{1/3} /(\ell^{2/3} \alpha)$, where $\nu_n$ is the hyperviscosity acting on large $k_\perp$, $\mu_m$ is the hyperviscosity acting on large $k_\parallel$ and $\alpha$ is the large-scale friction coefficient. 
 In the present framework, we are interested in monitoring the amplitude 
of the inverse cascade as a function of the parameter $\lambda$ in the limit of
large $\Rey_\nu,\Rey_\mu,\Rey_\alpha$ and large $\Lambda$.

Before we describe the simulations performed for this work, we define a few quantities of interest which we will use in the following. The 2-D energy spectrum is defined as
\begin{equation}
E(k_\perp,k_\parallel) = \frac{1}{2} \sum_{\mathbf{p}_\perp \atop k_\perp - \frac{1}{2} \leq p_\perp < k_\perp + \frac{1}{2}} \left( \frac{|\hat{\omega}_\parallel(\mathbf{p}_\perp,k_\parallel)|^2}{p_\perp^2} + |\hat{u}_\parallel(\mathbf{p}_\perp,k_\parallel)|^2\right), \label{eq:2Dspec}
\end{equation}
where hats denote Fourier transforms. The 1-D energy spectrum is obtained from (\ref{eq:2Dspec}) by summation over $k_\parallel$,
\begin{equation}
    E(k_\perp) = \sum_{k_\parallel} E(k_\perp,k_\parallel) \equiv E_\perp(k_\perp) +  E_\parallel(k_\perp) ,\label{eq:1Dspec}
\end{equation}
where $E_\perp$ contains all terms involving $\hat{\omega}_\parallel$ and $E_\parallel$ contains all terms involving $\hat{u}_\parallel$. The 2-D dissipation spectrum is defined as
\begin{equation}
D(k_\perp,k_\parallel) =\sum_{\mathbf{p}_\perp \atop k_\perp - \frac{1}{2} \leq p_\perp < k_\perp + \frac{1}{2}}\left[\left(\nu_n p_\perp^{2n} + \mu_m k_\parallel^{2m}\right)\left( \frac{|\hat{\omega}_\parallel(\mathbf{p}_\perp,k_\parallel)|^2}{p_\perp^2} + |\hat{u}_\parallel(\mathbf{p}_\perp,k_\parallel)|^2\right)\right]. \label{eq:2Ddspec}
\end{equation}
The large-scale energy dissipation rate is given by: 
\begin{equation}
    \epsilon_\alpha = \alpha \sum_{\mathbf{k},|\mathbf{k}_\perp|\leq 2} |\hat{u}(\mathbf{k})|^2 \label{eq:lsdiss}
\end{equation}
that measures the rate energy cascades inversely to the largest scales of the system.
Finally, the spectral energy flux in the perpendicular direction through a cylinder of radius $k_\perp$ in Fourier space is defined as
\begin{equation}
    \Pi(k_\perp) = \left\langle {(\mathbf{u})}_{k_\perp}^< \cdot [(\mathbf{\mathbf{u}}_\perp\cdot \nabla) \mathbf{u}]\right\rangle, \label{eq:flux}
\end{equation}
where $\mathbf{u}=(\mathbf{u}_\perp,{u}_\parallel)$, $\mathbf{u}_\perp = \hat{e}_\parallel \times \nabla \psi$ and 
\begin{equation}
    {(\mathbf{u})}_{k_\perp}^<  = \sum\limits_{\mathbf{p} \atop \text{ } p_\perp<k_\perp} \hat{\mathbf{u}} (\mathbf{p} )\exp(i \mathbf{p}\cdot\mathbf{x}) .
\end{equation}

\begin{table}                                                                         %
\begin{center}                                                                        %
   \begin{tabular}{ | c | c | c | c | c | c | c | c | c |  }                          %
\hline                                                                                %
    Set& A &B &C  \\
    $\Rey_\nu$ & $3.1\times10^3$ &$4.9\times10^5$ & $4.9\times10^5$   \\              %
    \NOTE{$\Rey_\mu$} & \NOTE{$1.9\times10^2$} & \NOTE{$1.9\times10^2$} & \NOTE{$1.9\times10^2$}   \\              %
     $\Lambda$ & 32 & 32 & 64 \\                                                          %
     resolution & $512^2\times n_z$&$1024^2\times n_z$&$1024^2\times n_z$ \\          %
     \# runs &  16 & 9 & 9                   \\ 
     \# $\tau_{eddy}$ & 22000 & 2500 & 7000
  \end{tabular}                                                                       %
    \caption{ Summary of the different simulations performed, where the resolution $n_z$ in the parallel direction 
              is varied between $128$ and $512$ in order to ensure well-resolved simulations.       %
              For each column, "\# runs" different values of $\lambda$, as defined in (\ref{eq:def_lamb}), were investigated, \NOTE{$\Rey_\mu$ lists the maximum value in each set} and $\#\tau_{eddy}$ gives the number of eddy turnover times $\tau_f=\epsilon_{in}^{-1/3} \ell^{2/3}$ simulated for each set of runs.}%
  \label{tab:runs}                                                                    %
\end{center}                                                                          %
\end{table}                                                                           %

The code used to solve equations (\ref{eq:red_vert_mom_dissmod}, \ref{eq:red_vert_vort_dissmod}) is based on the Geophysical High-order Suite for Turbulence, using pseudo-spectral methods including 2/3 aliasing to solve for the flow in the triply periodic domain, \citep[see][]{mininni2011hybrid}. We performed three sets of experiments, one at resolution $512^2\times n_z$ (set A) and two at $1024^2\times n_z$ (sets B and C), where the resolution $n_z$ in the parallel direction is varied depending on $\lambda$ from $128$ to $512$ to ensure well-resolvedness at minimum computational cost. We choose either $\Lambda=32$ (sets A and B) or $\Lambda=64$ (set C). The parameters $\nu_n$ and $\mu_m$ are chosen for every simulation so that  the run is well-resolved at large $k_\parallel,k_\perp$. This is checked by verifying that the maximum dissipation is captured within the interior of the 2-D dissipation spectrum (\ref{eq:2Ddspec}). The coefficient $\alpha$ was chosen so that that the 1-D spectrum (\ref{eq:1Dspec}) does not have a maximum at $k=1$ (i.e. no condensate is formed). We use random initial conditions whose small energy is spread out over a range of wavenumbers.
In each of the three sets of experiments, we keep $\Lambda$ and $\Rey_\nu$ fixed and vary $\lambda$ from small (less fast rotation, taller domain) to large (faster rotation, less tall domain). A summary is given in table \ref{tab:runs}.

In all simulations, we monitor the 1-D and 2-D energy spectra (\ref{eq:1Dspec}, \ref{eq:2Dspec}) as well as the large-scale dissipation rate (\ref{eq:lsdiss}). Simulations are continued until a steady state is reached where the large-scale dissipation rate and the energy spectrum are statistically steady, with the 1-D energy spectrum not having its maximum at $k=1$. Note that in such a steady-state situation $\epsilon_{in} = \epsilon_\alpha + \epsilon_{\nu,\mu}$, where $\epsilon_{\nu,\mu}=\sum_\mathbf{k} D(k_\perp,k_\parallel)$ is the dissipation rate due to hyperviscosity in the parallel and perpendicular directions, dominantly occurring at small scales. Monitoring $\epsilon_\alpha$ thus gives the amount of energy transferred inversely up to the largest scales $k=1,2$ and allows to measure the strength of the inverse cascade. Despite the fact that we solve asymptotically reduced equations, which allows larger time steps, the required simulation time was non-negligible since convergence to the steady state was slow in some cases. In total, more than 30000 forcing-scale-based eddy turnover times $\tau_f = \epsilon^{-1/3} \ell^{2/3}$ were simulated, amounting to around two million CPU hours of computation time.

\section{Results from direct numerical simulations}        
\label{sec:results}                                        
\subsection{Transition to an inverse cascade}  
\label{sec:results1}                           
\begin{figure}
    \centering
    \includegraphics[width=6.5cm]{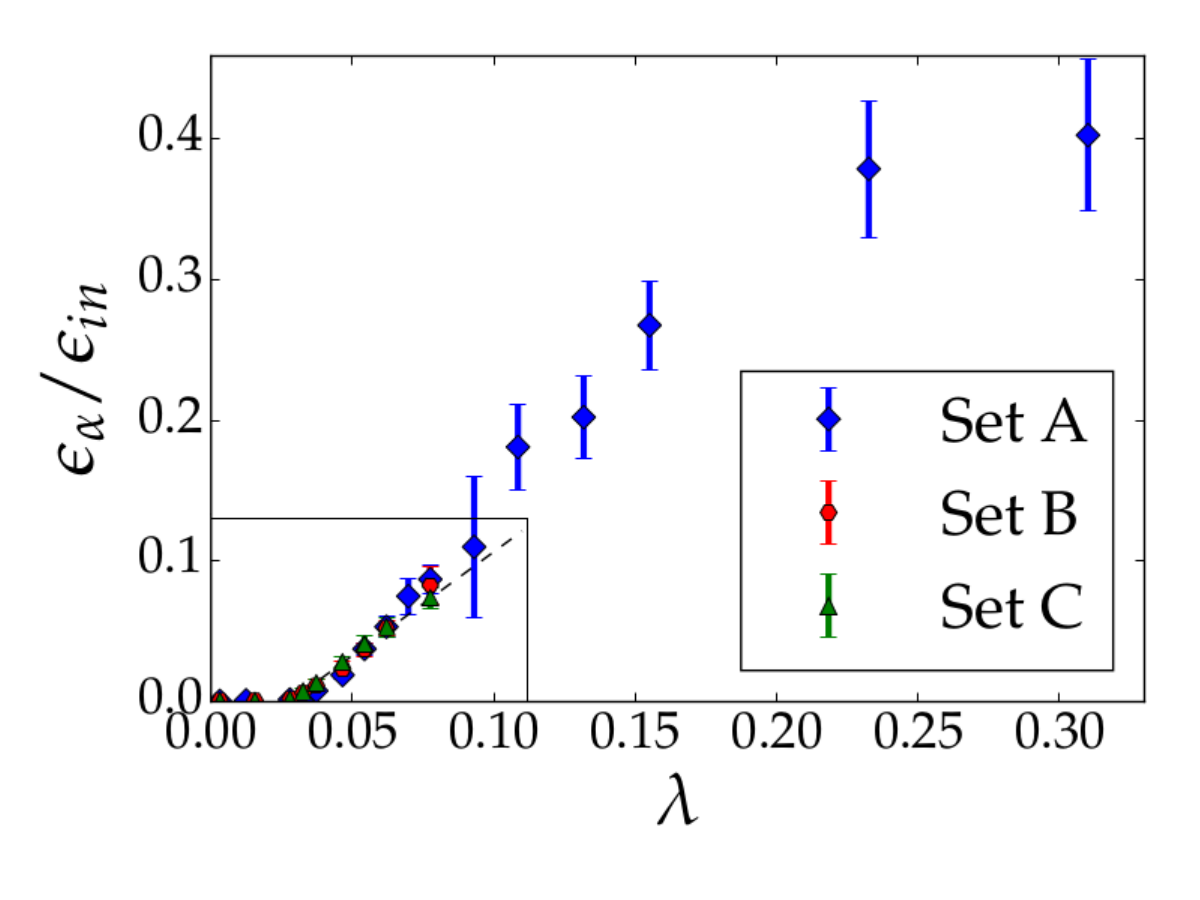} \includegraphics[width=6.5cm]{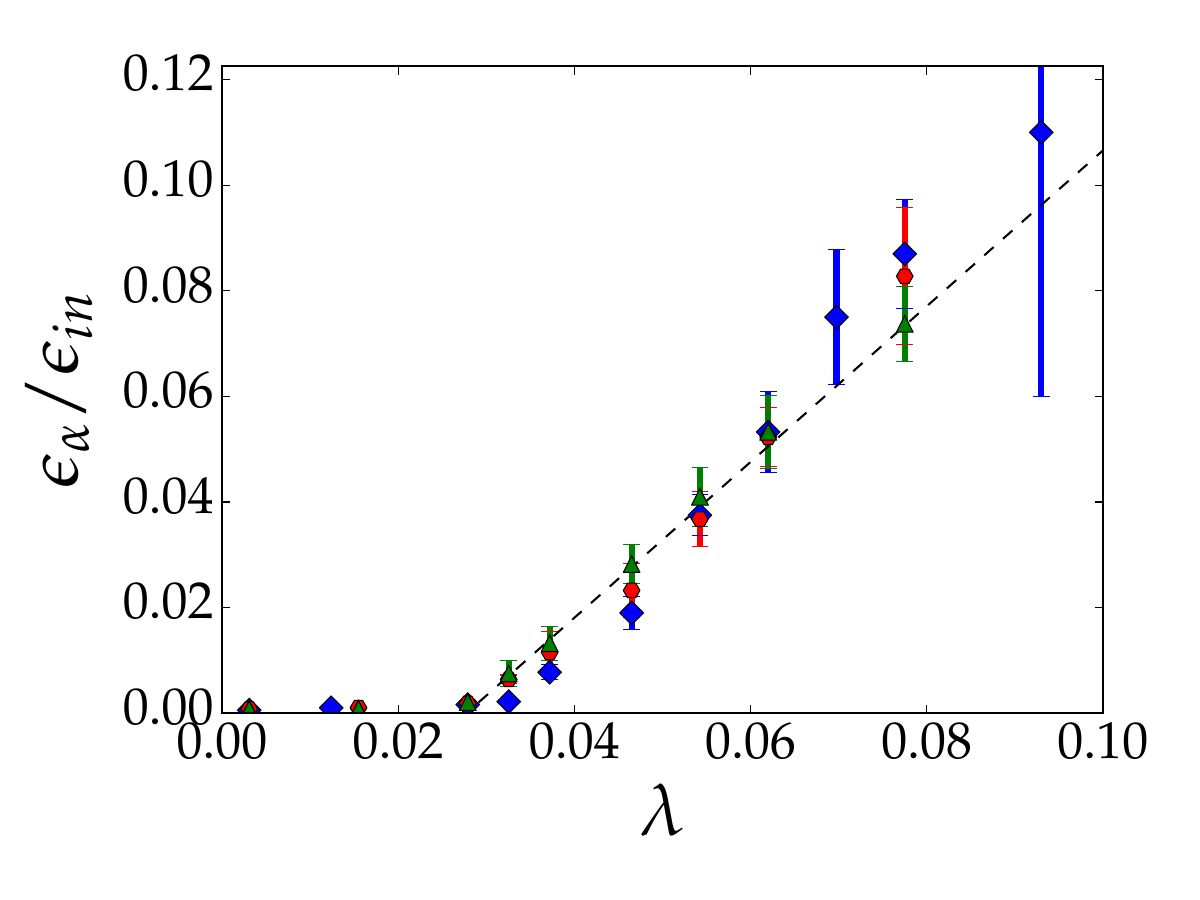}
    \caption{Large-scale dissipation rates as defined by (\ref{eq:lsdiss}) measured in steady state from sets A, B and C for different values of $\lambda$. Error bars correspond to standard deviation in steady state. The black dashed line is a linear fit based on set C. Right: all values of $\lambda$, Left: zoom close to $\lambda=\lambda_c$.}
    \label{fig:eps_inv_vs_lamb}
\end{figure}

In this section, we present the results of the direct numerical simulations (DNS) obtained in steady state. 
 The central goal of this work is to determine the properties of the transition 
from a strictly forward cascade to a state with an inverse cascade. 
 The amplitude of the inverse cascade is given by the large-scale dissipation rate $\epsilon_\alpha$ that measures the rate at which energy is transferred to the large scales. In the presence of an inverse cascade, $\epsilon_\alpha$ converges to a finite value in the limit of $\Lambda ,\Rey_\alpha,\Rey_\mu,\Rey_\nu\to \infty$, while it converges to zero in the absence of an inverse cascade. In Figure \ref{fig:eps_inv_vs_lamb} we show  $\epsilon_\alpha$ (time averaged at steady-state) as a function of the parameter $\lambda$ from all simulations. One observes a transition from  $\epsilon_\alpha/\epsilon_{in}\approx 0$ to finite values at $\lambda=\lambda_c \approx 0.03$. At $\lambda<\lambda_c$ no inverse cascade is present and a vanishingly small amount energy reaches the scales $k_\perp=1,2$, where the large-scale dissipation acts. However, for $\lambda>\lambda_c$ an inverse cascade develops, whose strength increases monotonically with $\lambda-\lambda_c$, leading to non-vanishing large-scale dissipation. Comparing the curves obtained from sets $A$, $B$ ($\Rey_\nu$ increased) and $C$ ($\Rey_\nu$ and horizontal box size $\Lambda$ increased), one observes that the transition appears to become sharper with increasing Reynolds number and box size, and remains at the same point. This indicates that the transition is likely to be critical and continuous, having a discontinuous 1st derivative at $\lambda_c$ in the limit $\Rey_\nu,\Lambda\to \infty$. Considering only the highest $\Rey_\nu$ and $\Lambda$, i.e. set C only, we estimate from figure \ref{fig:eps_inv_vs_lamb} that $\epsilon_\alpha \propto (\lambda-\lambda_c)^\gamma$ with $\gamma\approx 1$ from a fit close to onset, within our uncertainties. However, this estimate of the critical exponent is not definitive and a larger number of simulations and parameter values are needed to ascertain its precise value with higher confidence.
\begin{figure}
    \centering
    \includegraphics[width=0.475\textwidth]{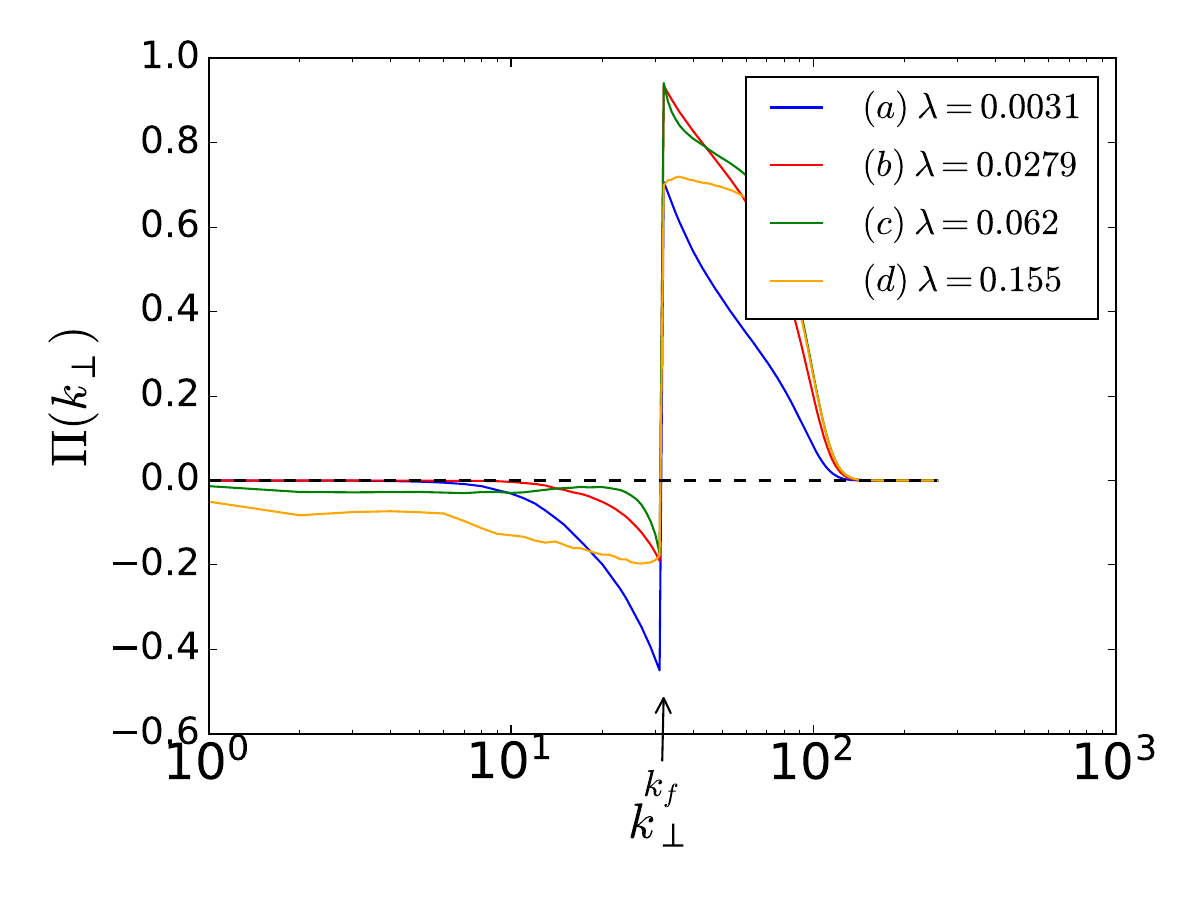}
    \includegraphics[width=0.475\textwidth]{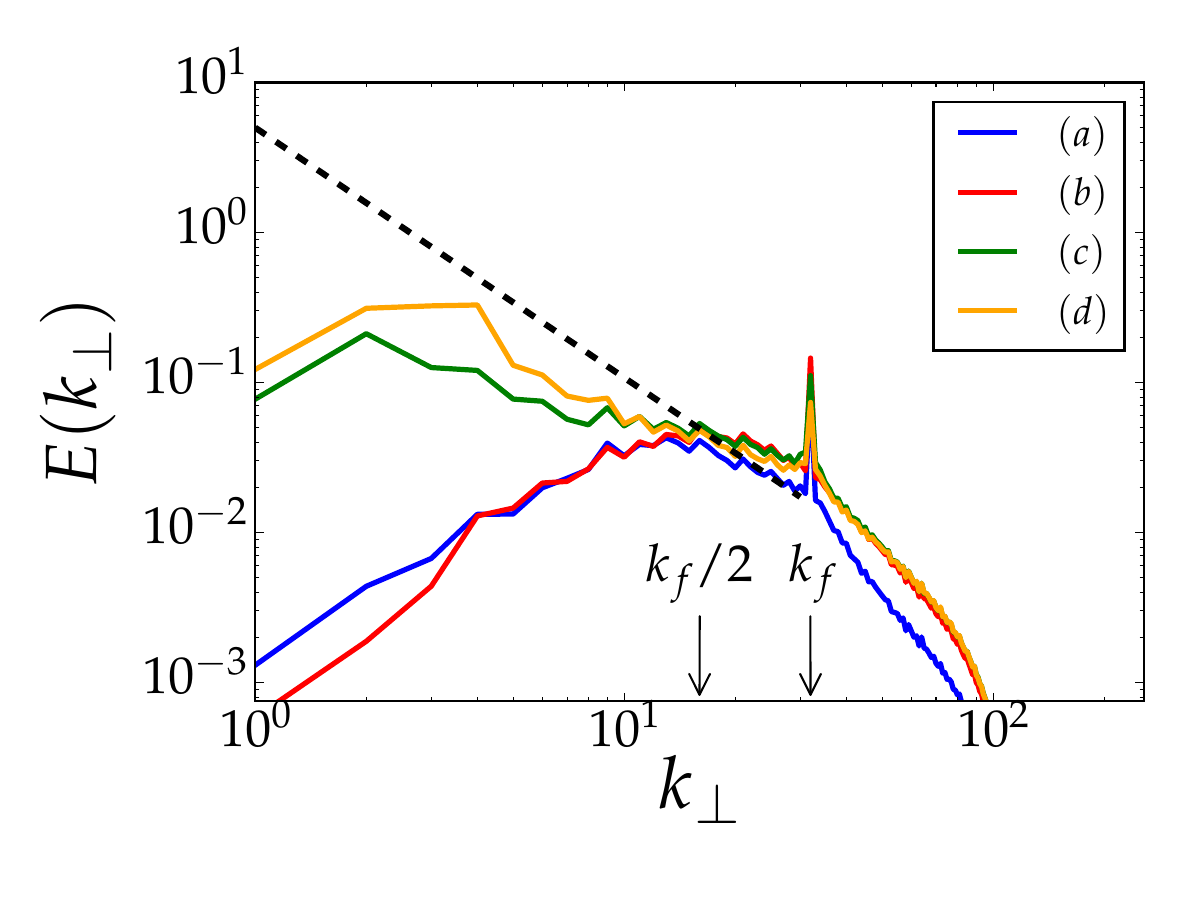}
    \caption{Left: Steady-state energy flux in the perpendicular direction as a function of perpendicular wavenumber for four different values of $\lambda$ from set A. Right: Corresponding steady-state 1-D energy spectra. }
    \label{fig:fluxes}
\end{figure}
 The left panel of figure \ref{fig:fluxes} shows the energy flux in steady state for four values of $\lambda$ from set A, namely $(a)$ $\lambda = 0.0031$, $(b)$ $\lambda = 0.00279$, $(c)$ $\lambda = 0.062$ and $(d)$ $\lambda=0.155$. 
 Cases $(a)$ and $(b)$ correspond to $\lambda<\lambda_c$, while for cases $(c)$ and $(d)$ $\lambda>\lambda_c$. 
 All simulations present a significant forward energy flux for $k>k_f$. 
 For $k\ll k_f$ the energy flux vanishes for the small-$\lambda$ cases (a) \& (b) \NOTE{(lower rotation rates, taller boxes)}. Some inverse flux is observed for these cases, which is however confined to around $k\approx k_f/2$. By contrast, a non-vanishing inverse energy flux extending up to $k_\perp=1$ is observed for the larger-$\lambda$ cases (c) \& (d) (higher rotation rate, shallower box) that display an inverse cascade.  

\subsection{Energy Spectra}                               
\label{sec:results2}                                      

The right panel of \ref{fig:fluxes} shows the corresponding 1-D spectra for the same four values of $\lambda$ as in the left panel of the same figure.
In cases $(c)$ and $(d)$, that display an inverse cascade, the spectrum is maximum at small wave numbers $k_\perp\simeq 2$.
The reason why the spectrum does not peak at the smallest wavenumber $k=1$ is the damping by the large-scale friction. 
In cases $(a)$ and $(b)$, the spectrum has two local maxima, one at the forcing scale $k_\perp=k_f$ and another one near $k_\perp=k_f/2$. This implies that there is transfer of energy to scales twice as large as the forcing scale. This, however, does not indicate an inverse cascade as this secondary peak remains close to the forcing scale and does not move further up to larger scales.

\begin{figure}
    \centering
    \includegraphics[width=0.475\textwidth]{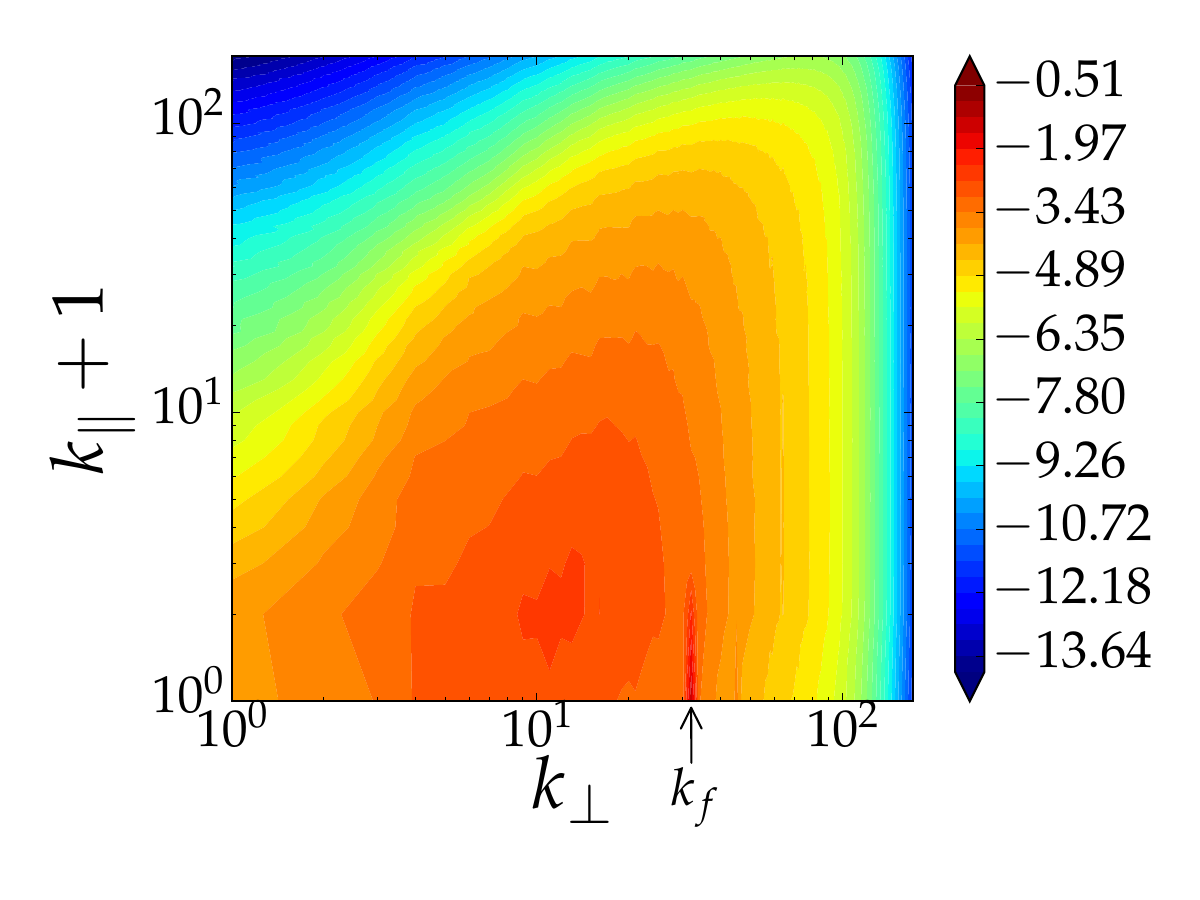}
    \includegraphics[width=0.475\textwidth]{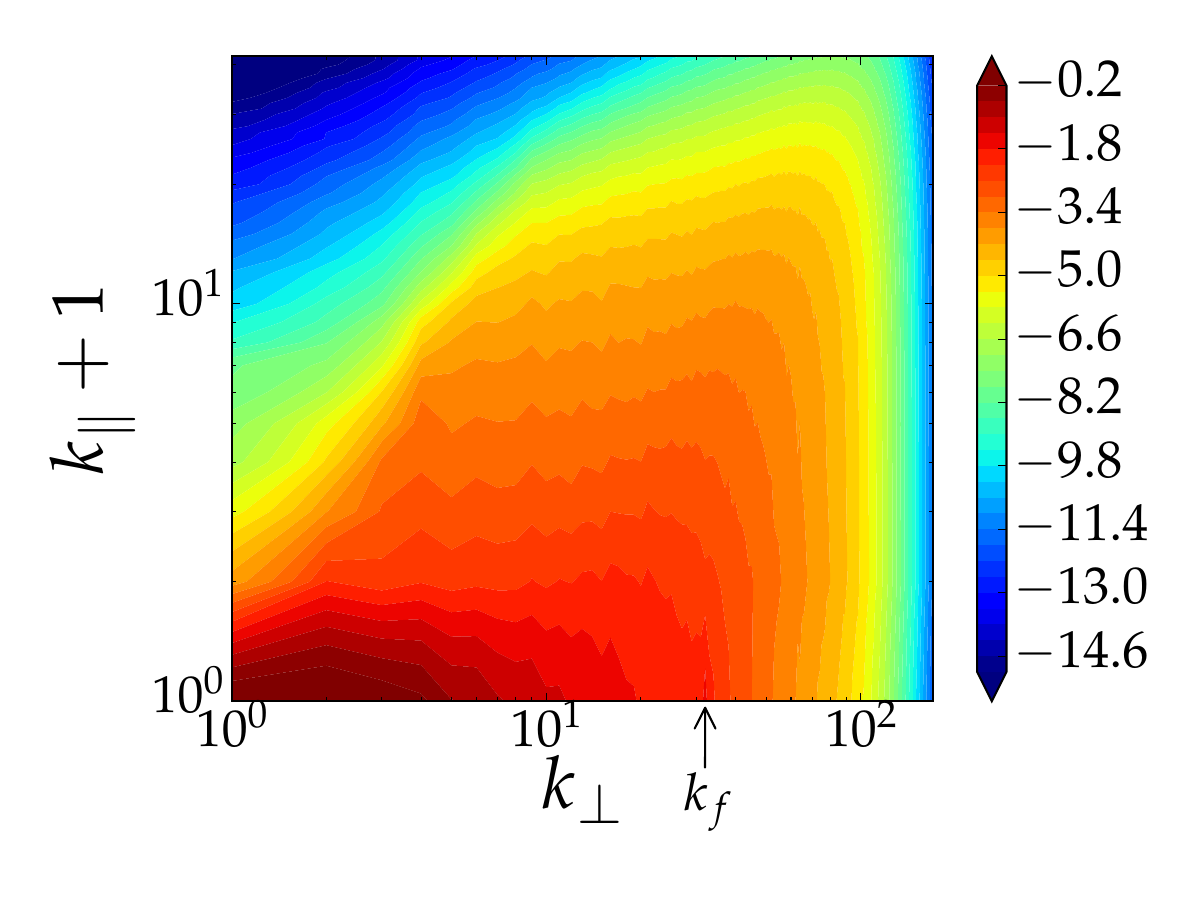}
    \caption{Steady-state 2-D energy spectra as defined in (\ref{eq:2Dspec}) from set A for $\lambda=0.0279$ (left) and $\lambda=0.155$ (right). Color bar logarithmic with base $10$.}
    \label{fig:2Dspec}
\end{figure}

\begin{figure}
    \centering
    \includegraphics[width=0.475\textwidth]{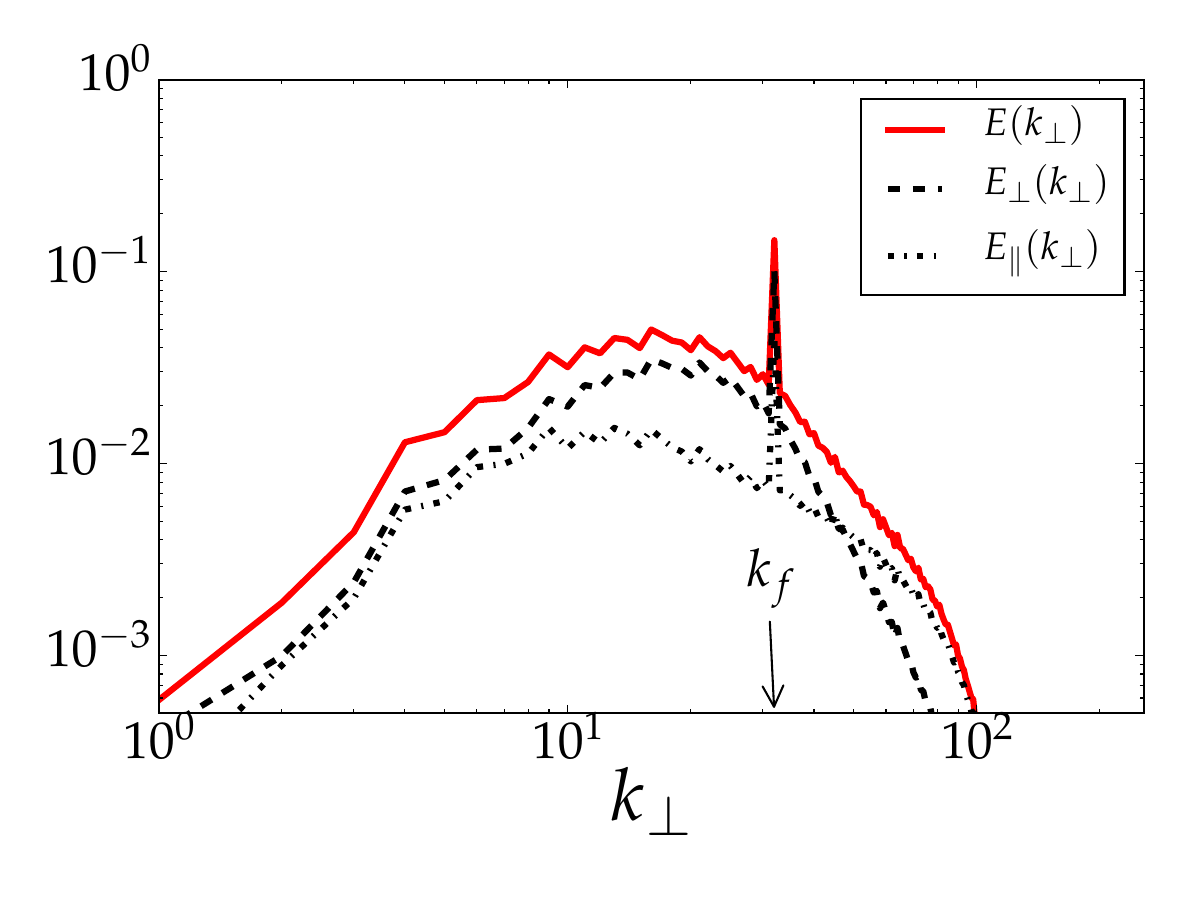}
    \includegraphics[width=0.475\textwidth]{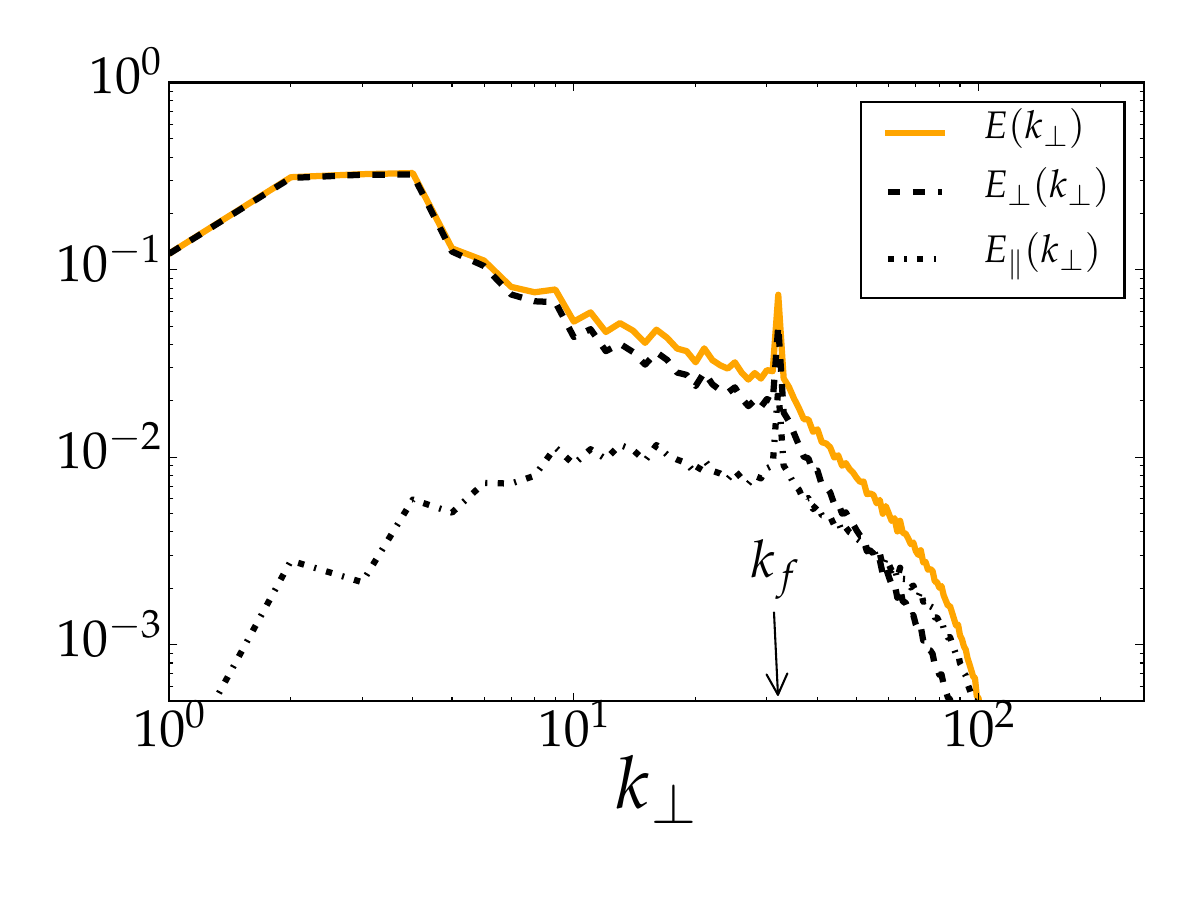}
    \caption{Left: 
    Decomposition of spectrum at $\lambda=0.0279$ into contributions $E_\perp$ from perpendicular motions and $E_\parallel$ from parallel motions. 
    Right: same as left for $\lambda=0.155$.}
    \label{fig:1Dspec}
\end{figure}

The 2-D spectra associated with cases $(b)$ and $(d)$ are presented in figure \ref{fig:2Dspec}. They show that the secondary maximum observed in the 1-D energy spectra at $k_\perp \approx k_f/2$ for $(b)$ stems from contributions at $k_\parallel>0$. 
For $\lambda>\lambda_c$, the inverse energy cascade of the 2-D manifold leads to a maximum at $k_\parallel =0$, at small $k_\perp$. 
%
Finally, figure \ref{fig:1Dspec} shows the 1-D spectra from cases  $(b)$ and $(d)$ decomposed to their perpendicular $E_\perp(k_\perp)$ and parallel $E_\parallel(k_\perp)$ components. They show that perpendicular motions dominate for all wavenumbers $k<k_f$ in the case of an inverse cascade and also close to the secondary maximum at $k_f/2$ for the flows that do not display an inverse cascade. At large $k>k_f$, the two spectra are of the same order with $E_\parallel(k_\perp)> E_\perp(k_\perp)$. \NOTE{One further observes that when an inverse cascade is present, it is occurring in the perpendicular components only. This is in agreement with expectation, since the parallel velocity component obeys an advection-diffusion equation in the slow manifold, and therefore displays a forward cascade.} 
 %

The peak observed in the 1-D spectrum at $k\approx k_f/2$ (which occurs here for all cases that do not display an inverse cascade) is unexpected and deserves some further discussion. First we should note that this is not the first time a similar feature is observed. In \cite{buzzicotti2018inverse}, where simulations of rotating turbulence were performed, artificially excluding the $k_\parallel=0$ plane in Fourier space showed a similar maximum. \NOTE{More recently such a maximum was also observed in simulations of rotating turbulence in elongated domains \cite{DiClarc_to_appear}.}
%
%
Since this is the statistically steady state of the system and energy does not cascade further upscale, this inverse transfer does not stem from a turbulent inverse cascade, which would continue up to the largest scales, as it does for $\lambda<\lambda_c$. We have also verified that starting from  initial conditions obtained from a run with $\lambda > \lambda_c$ and decreasing $\lambda$ to a value below $\lambda_c$ resulted, at long times, in a state with no inverse cascade. 
Rather, one may suspect an instability mechanism involving the forcing-scale flow.

\NOTE{Indeed, the linear stability analysis presented in section \ref{subsec:homoch_instab}, considering a homochiral wavenumber triad comprising one large-amplitude mode $\mathbf{k}=(k_f,0,0)$ at the forcing scale and two small-amplitude inertial waves at $\mathbf{p},\mathbf{q}$, gives that the inertial waves with $q_\perp\approx k_f/2$, and $|q_\parallel|\lesssim 0.35 k_f/\lambda$ are linearly unstable. Interestingly, \citep{buzzicotti2018inverse} also found homochiral interactions to be responsible for the inverse energy transfer in their simulations. This instability can explain in part the transfer of energy
to the $k_f/2$ modes. We note, however, that the maximum growth occurs at $q_\parallel=0$ for the triad (see figure \ref{fig:growth_rate_triad}), which is not where the maximum is observed in the 2-D spectra shown in figure \ref{fig:2Dspec}. }

\begin{figure}
    \centering
    \includegraphics[width=6.0cm]{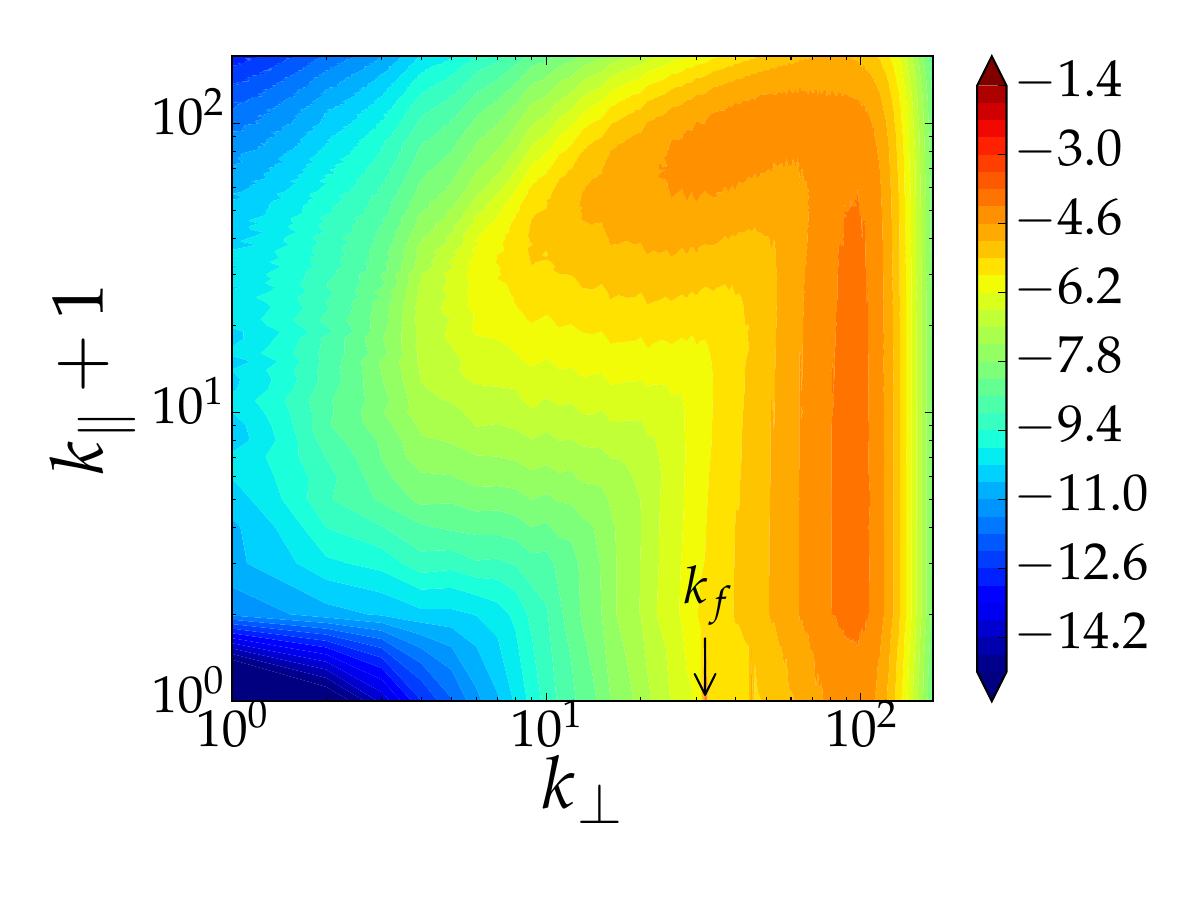}
    \includegraphics[width=6.0cm]{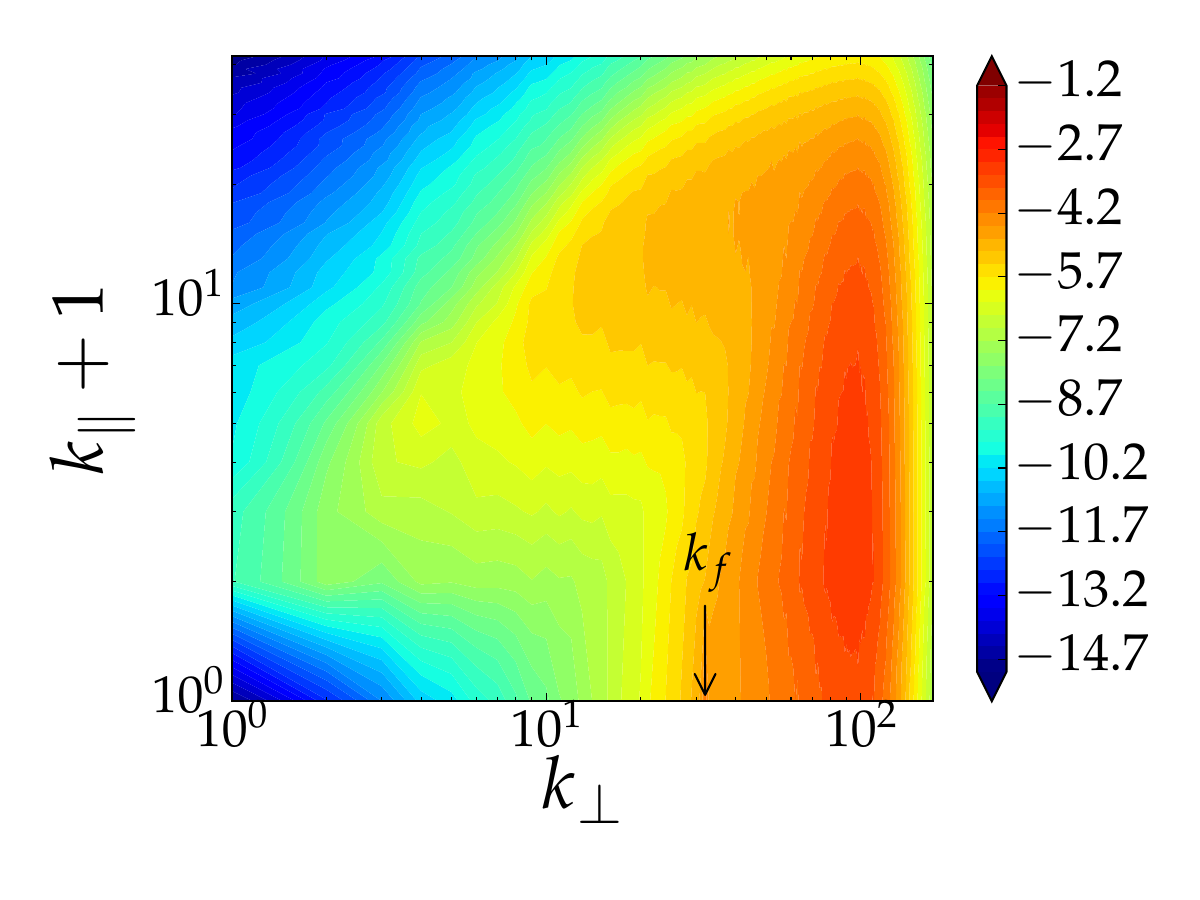}
    \caption{Steady-state 2-D dissipation spectra from set A for $\lambda=0.0279<\lambda_c$ (left) and $\lambda=0.155>\lambda_c$ (right). Color bars are logarithmic with base $10$. }
    \label{fig:2Ddspec}
\end{figure}
\begin{figure}
    \centering
    \includegraphics[width=6cm]{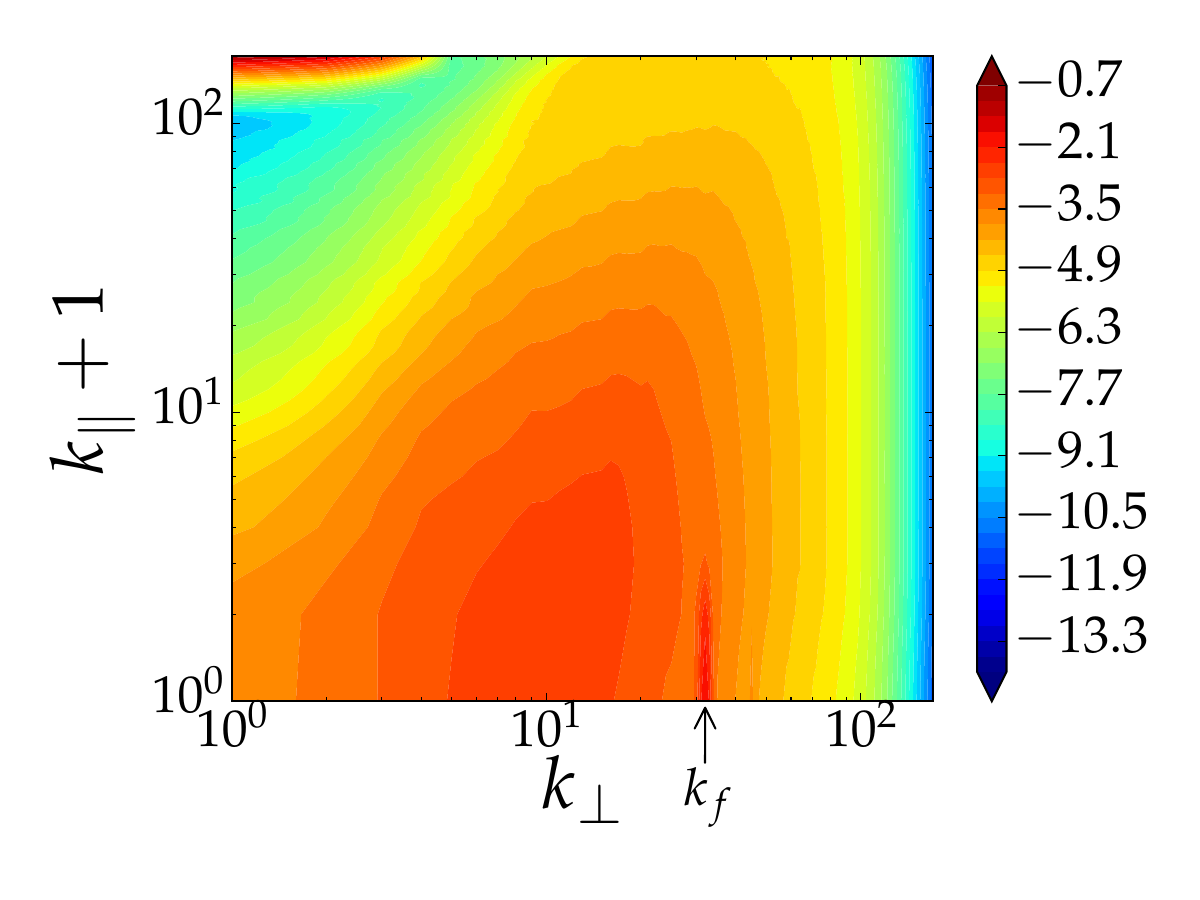}
    \includegraphics[width=6cm]{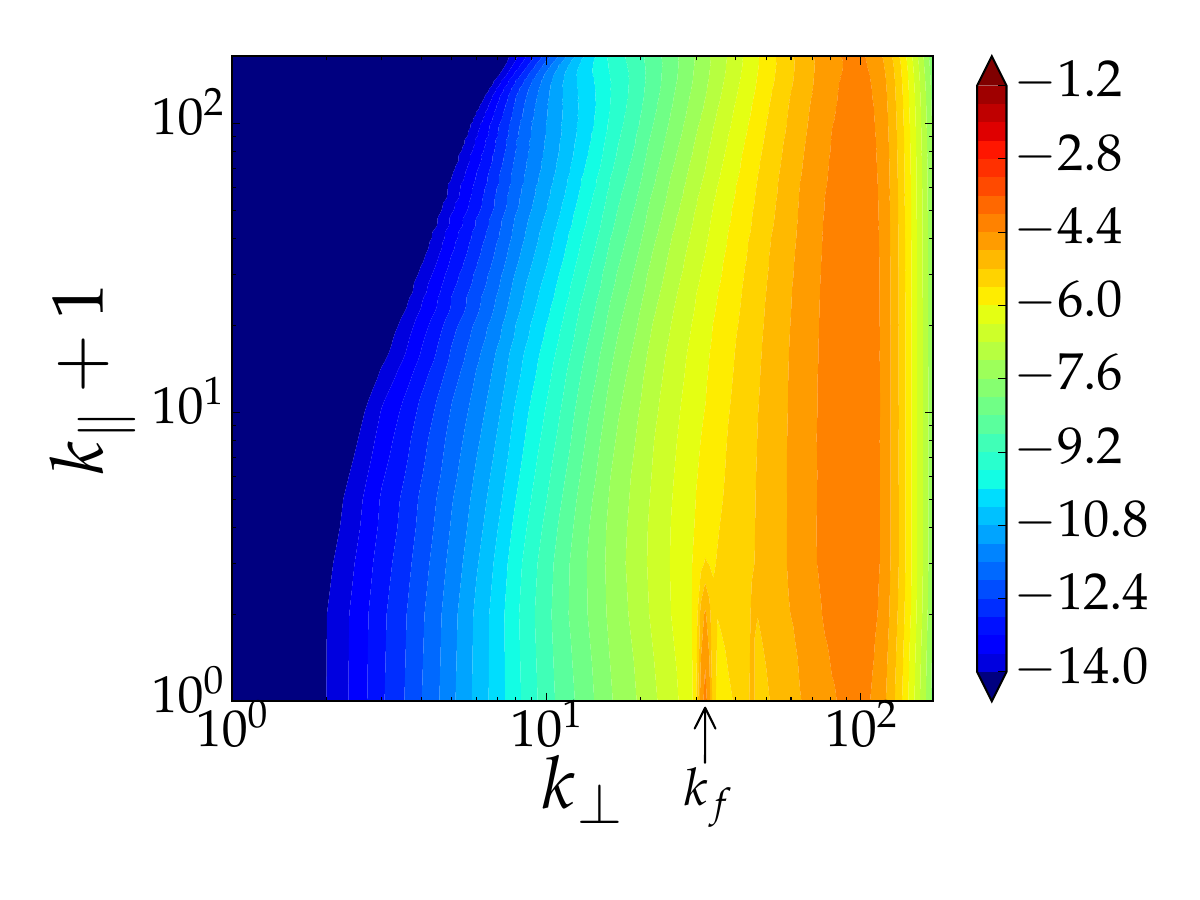}
    \caption{\NOTE{Left: 2-D dissipation spectrum from a run at $\Rey_\mu=\infty$ from set A, with $\lambda=0.01<\lambda_c$. A maximum at large $k_\parallel$ forms, which is absent at $\Rey_\mu<\infty$. Right: 2-D dissipation spectrum from a run at $\Rey_\mu=\infty$ from set A, with $\lambda=0.01<\lambda_c$. By contrast with the runs at $\Rey_\mu<\infty$, the dissipation spectrum tends to become independent along $k_\parallel$. Color bars are logarithmic with base $10$.} }
    \label{fig:2Ddspec_mu0}
\end{figure}

\subsection{Convergence}                       
\label{sec:results3}                           
%
In figure \ref{fig:2Ddspec} we also show the 2-D energy dissipation spectra for the same cases as in figure \ref{fig:2Dspec}. The dissipation spectra demonstrate the well-resolvedness of the simulations:
the maximum dissipation is within the simulation domain and not at the maximum wavenumbers 
(in the parallel or perpendicular direction).  It is worth noting that most of the dissipation
is occurring at large $k_\perp$ and not at large $k_\parallel$. 
The artificial hyper-viscosity used for the parallel wavenumbers in the simulations thus plays a minor role in dissipating energy. This is important because in the asymptotic expansion (eqs. \ref{eq:red_vert_mom},\ref{eq:red_vert_vort}), only the perpendicular wavenumbers participate in the dissipation. 

\NOTE{The reason why we have nonetheless added an artificially finite $\Rey_\mu$ becomes apparent in figure \ref{fig:2Ddspec_mu0} where results from a simulation without the dissipation at large $k_\parallel$ are shown. In this case a spurious maximum forms in the 2-D energy spectrum at large $k_\parallel$ and small $k_\perp$, and the dissipation spectrum tends to become invariant along $k_\parallel$. This implies a violation of the criterion for well-resolvedness, since the maxima of the energy and dissipation spectra are not contained within the domain, but touch the domain limit at large $k_\parallel$. This side effect can be circumvented by increasing the resolution the parallel direction significantly, but this would  increase the numerical cost of the study and has therefore been avoided.}
It is also worth noting that for case $(b)$ ($\lambda=0.0279$) a higher resolution in the parallel direction was
required than for case (d) ($\lambda=0.155$). This is because the small-$\lambda$ flows are more efficient 
at generating small scales in the parallel direction, while such generation is suppressed in large-$\lambda$ flows by rotation.

\subsection{Spatial Structures}                           
\label{sec:results4}                                      

\begin{figure}
    \centering
    \includegraphics[width=7.25cm]{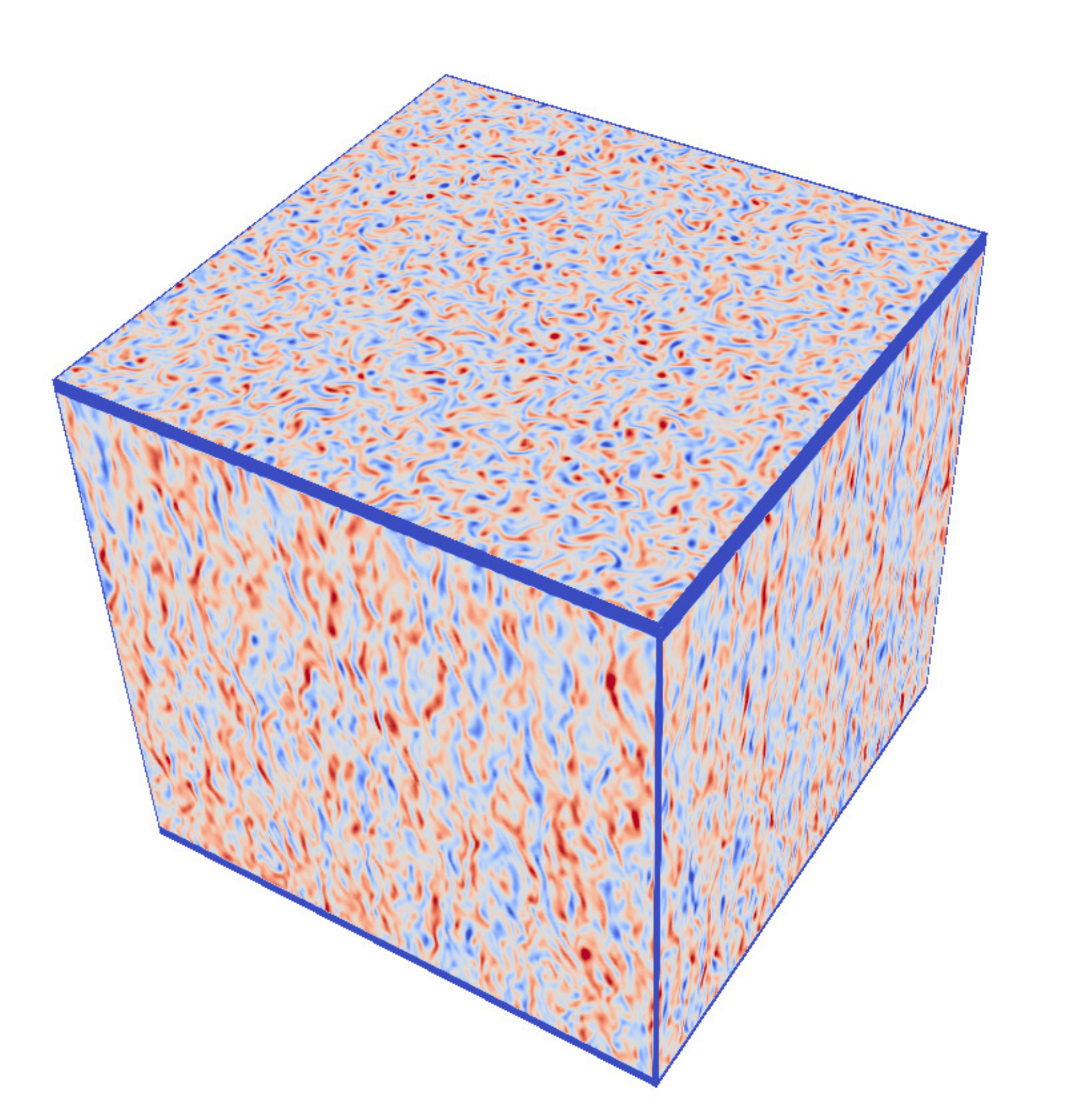}\includegraphics[width=7.25cm]{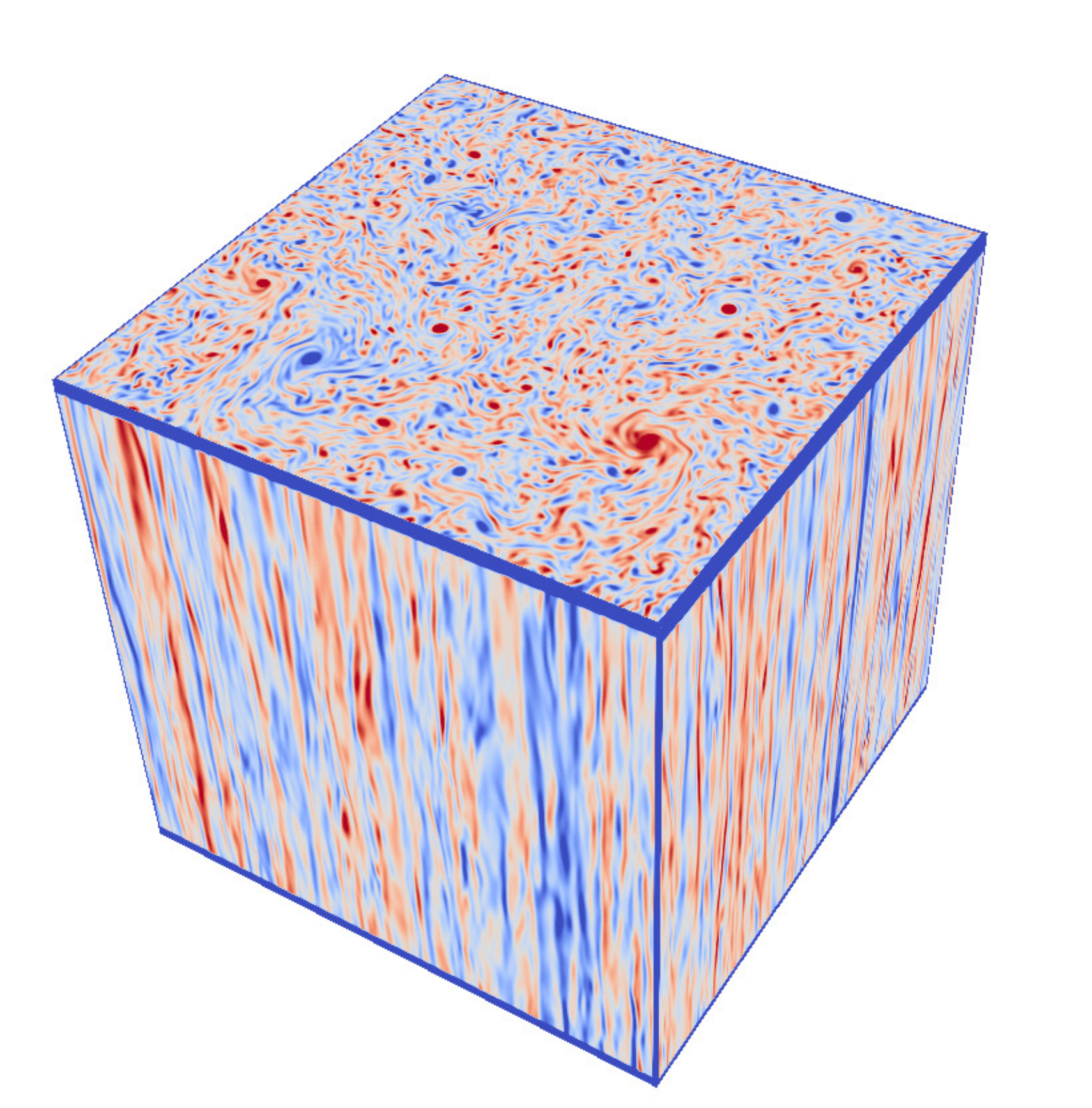}
    \caption{Visualisation, \NOTE{ in the asymptotic scaling}, of the flow in terms of vorticity $\omega_\parallel$ at $\lambda = 0.027 \lesssim \lambda_c$ (left) and $\lambda=0.23>\lambda_c$ (right). Positive vorticity in red, negative in blue, the edges have been coloured blue for better visualisation. For $\lambda>\lambda_c$, on the sides of the domain one can see elongated structures along the parallel direction, while the on top of the domain well-separated vortices are seen. For $\lambda<\lambda_c$, these elongated structures are absent. Furthermore, the flow for $\lambda>\lambda_c$ is characterised by larger perpendicular scales than the flow at $\lambda<\lambda_c$.} 
    \label{fig:visualisation}
\end{figure}

\begin{figure}
    \centering
    \includegraphics[width=7.25cm]{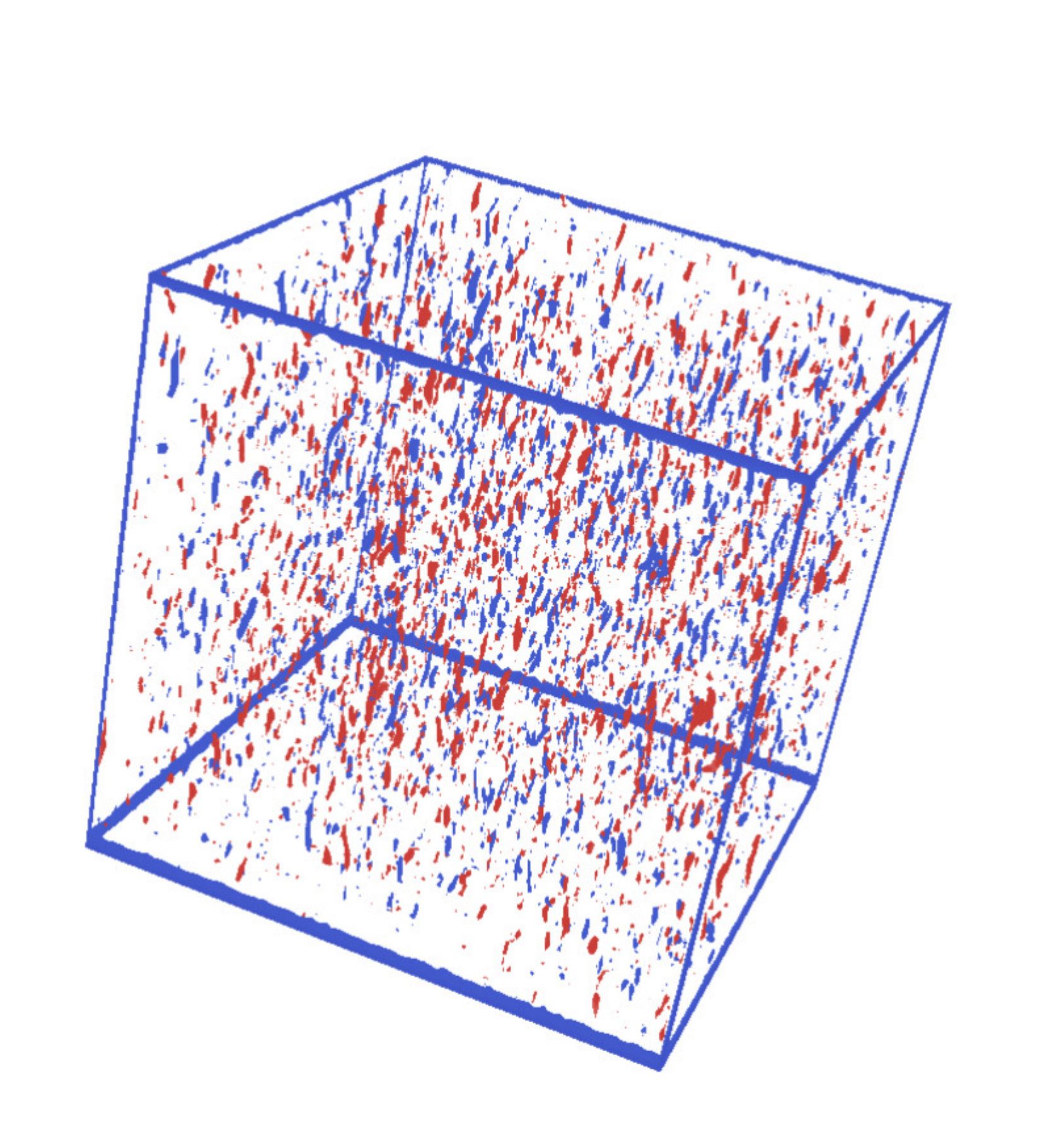}\includegraphics[width=7.25cm]{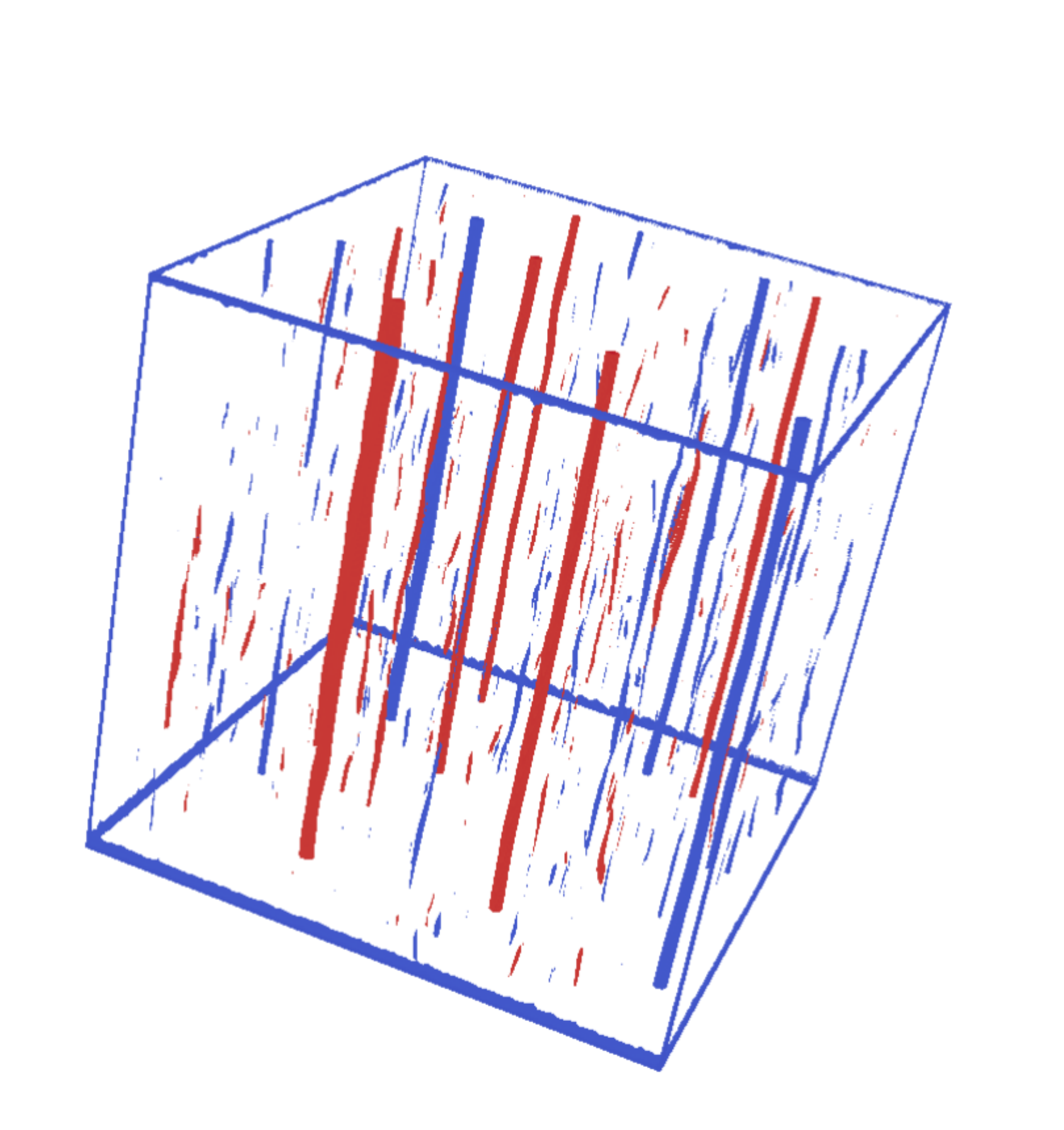}
    \caption{\FINALNOTE{Same as in figure \ref{fig:visualisation}, but with reduced opacity and filtered in vorticity, showing the most intense vortical structures only.}} 
    \label{fig:visualisation_opred}
\end{figure}

Finally, figure \ref{fig:visualisation} shows a visualisation of the flow in terms of vorticity $\omega_\parallel$ at $\lambda = 0.027 \lesssim \lambda_c$ (left) and $\lambda=0.23>\lambda_c$ (right).  \FINALNOTE{The same fields are shown once more with a reduced opacity in figure  \ref{fig:visualisation_opred}.} For $\lambda>\lambda_c$, columnar vortices are clearly visible which are approximately invariant along the axis of rotation. In the perpendicular direction these vortices are visibly of larger scale and organised in clusters. For $\lambda<\lambda_c$, no such anisotropic organisation of the flow can be observed.

\section{Conclusions}             
\label{sec:conclusions}           

\begin{figure}
    \centering
    \includegraphics[width=0.75\textwidth]{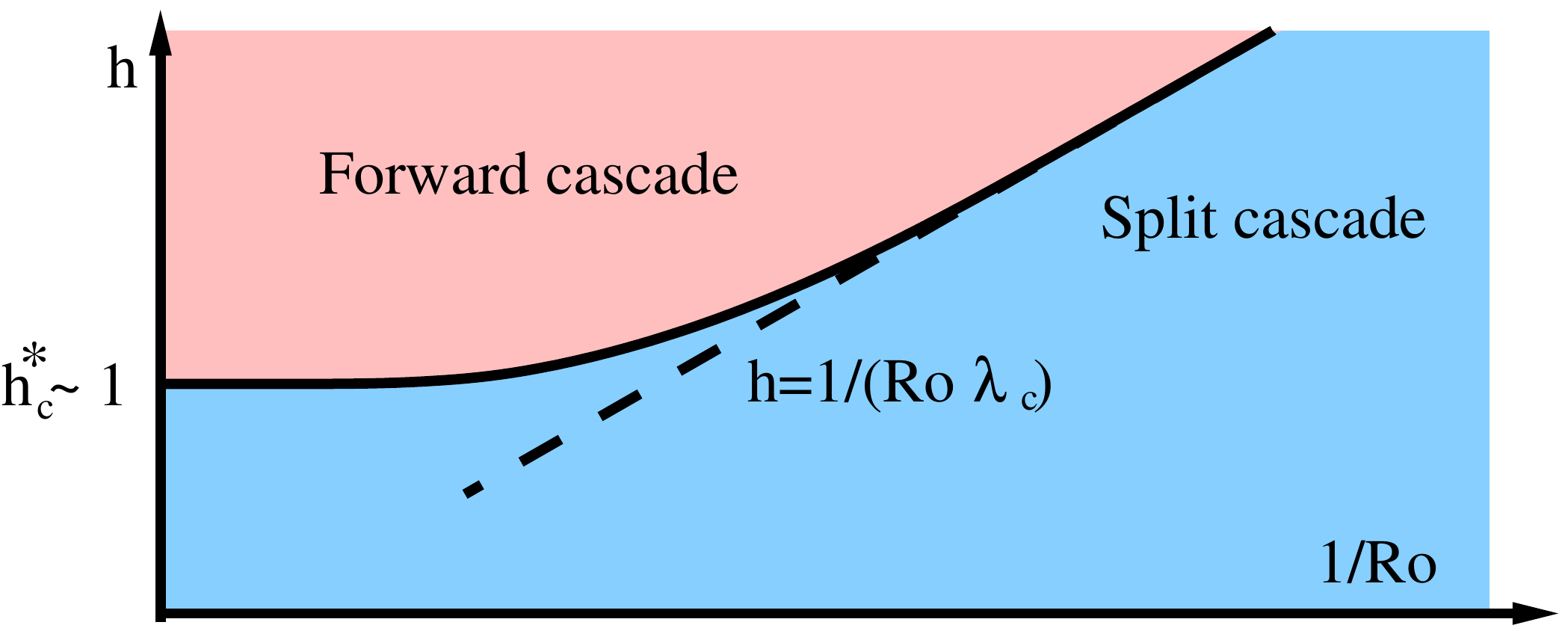}
    \caption{Phase space of rotating turbulence. The solid curved line indicates the critical line between forward and split cascade phases. The dashed straight line represents the major result obtained here: the asymptote of the critical line at large $1/Ro$ and large $h$. } 
    \label{fig:phase_space}
\end{figure}

%
\NOTE{In this work we investigated fast-rotating turbulence in elongated domains using an asymptotic expansion. 
A linear stability calculation of a single triad of homochiral modes of our model predicted that there is a critical value of 
the control parameter $\lambda$ below which the 3-D modes ($q_\parallel\ne0$) become unstable. Based on the fact that the 3-D modes favour a forward cascade, while the 2-D modes ($q_\parallel=0$) favour an inverse cascade, a transition of the cascade direction was anticipated. Indeed, the numerical simulations presented in section \ref{sec:results} indicated that there is a transition from a strictly forward cascade to a split cascade (where part of the energy cascades inversely) as the parameter $\lambda$ given in (\ref{eq:def_lamb}) is varied.
Since $\lambda$ is the only control parameter appearing in the reduced equations (\ref{eq:red_vert_mom_dissmod},  \ref{eq:red_vert_vort_dissmod}) that remains finite in the limits of infinite Reynolds number and infinite domain size, it uniquely determines the transition in the examined limit.
}
This result implies that if the limit of infinite domain height $h\to \infty$ is taken for fixed $Ro$, 
then $\lambda \to0$ and energy  cascades forward. On the other hand, if the limit $Ro\to0$ is taken
for a fixed domain height, then $\lambda \to \infty$ and an inverse cascade will be present. The fact that a transition to an inverse cascade is observed in the asymptotic limit $h\propto \Ro^{-1}\to \infty$, which is considered here numerically, confirms the theoretical arguments presented in section \ref{sec:theo_bg}. The phase space of rotating turbulence in the $(h,1/Ro$) plane, based on the present results, is as depicted in figure \ref{fig:phase_space}. In the limit of infinite $Re$ and $\Lambda$ two phases exist, one where there is only a forward cascade and one where there is a split cascade. They are separated by a critical line $h_c(Ro)$ that approaches the known non-rotating critical height $h^*_c$ for $\Ro\to\infty$, while for small $Ro$, which is the limit examined in the present work, $h_c$ scales like  $h_c=1/(Ro\lambda_c)$ with $\lambda_c \simeq 0.03$.

\NOTE{It is worth noting that weak wave turbulence is not met in our system. 
This is because in our expansion the limit $h\to\infty$ is taken together with the limit $\Ro\to 0$, keeping $\lambda$ fixed, while for weak wave turbulence one must take $h\to\infty$ first and then $Ro\to 0$ 
\citep{nazarenko2011wave}. Even when the limit $\lambda\to \infty$ is taken in our reduced equations, one does not recover weak wave turbulence but rather 2-D (strong) turbulence \citep{gallet2015exact}. On the other hand, when the limit $\lambda\to 0$ is taken we do obtain wave turbulence that cascades energy forward, but which is not weak since $\lambda \ll1$ implies that the wave periods are of the same order or longer than the eddy turnover time. Turbulence in our system is thus always strong.}

Our approach was based on asymptotic reduction, allowing us to reliably achieve the extreme parameter regimes required to test the theoretical predictions at comparatively moderate numerical cost. \NOTE{The asymptotics are valid under the conditions $\Ro \ll 1$, $\Rey \ll \Ro^{-1}$, $\Rey \ll H/(\ell_{in})$ and $L\ll H$. The efficiency of numerical integration of the reduced equations is due to the filtering of fast inertial waves by the Taylor-Proudman constraint.} We stress, however, that a question of order of limits arises. By solving the asymptotically reduced equations and investigating increasing $\Rey_\nu$ and $\Lambda$ in that framework, we are taking limits in the order 
\begin{equation}
      \lim  \limits_{\mathrm{Re}_\nu\to \infty   \text{ } \atop \text{ } \Lambda \to \infty}
\left(\lim  \limits_{{\Ro\to 0 \text{ }            \atop \text{ } \lambda=cst.}}
 \epsilon_\alpha (\lambda) \right) 
\quad \mathrm{as\,\, opposed\,\, to} \quad
      \lim  \limits_{{\Ro\to 0 \text{ }            \atop \text{ } \lambda=cst.}}
\left(\lim  \limits_{\mathrm{Re}_\nu\to \infty   \text{ } \atop \text{ } \Lambda \to \infty}
 \epsilon_\alpha (\lambda) \right), \label{eq:order_limits}
\end{equation}
{\it i.e.} first the low-Rossby-number limit (at fixed $\lambda$) is taken and then the large-Reynolds-number limit, and not vice versa, which would correspond to studying the ($\Ro$,$h$) dependence of an already fully turbulent flow. A priori, the two limits do not necessarily commute and therefore it is important to additionally study turbulent flows at finite rotations and domain heights $h$ in the full rotating Navier-Stokes system. \NOTE{This has recently been examined by \cite{DiClarc_to_appear}, where a new 
meta-stable {\it vortex crystal } state was found near the transition, in which co-rotating vortices organised themselves in a crystal. Such vortex-crystal states did not appear in the asymptotic model investigated here. Possibly, the true phase space of rotating turbulence can thus be considerably more complex. }

 Our numerical evidence also suggests that this transition is continuous but not smooth. The inverse cascade starts at a critical value $\lambda_c$ with an almost linear dependence on the deviation from criticality $\epsilon_\alpha\propto (\lambda-\lambda_c)$. Despite the simplicity of this behaviour, its origin is far from being understood. Similar scaling behavior has been found for the transition to the inverse energy cascade in thin-layer turbulence below a critical layer height $H_c$ \citep{benavides_alexakis_2017} and for the two-dimensional magnetohydrodynamic flow studied in \citep{seshasayanan2016critical,seshasayanan2014edge}. In both cases, a critical exponent close to unity is identified for the inverse energy transfer rate close to the transition. Future research should aim to provide an understanding of the origin of this estimated critical exponent. It should also verify which other turbulent fluid flows present criticality at the transition to the inversely cascading regime and whether their critical exponent is identical to or different from unity. Experimental studies of such systems, where long-time averages can be performed, may prove invaluable in understanding this non-equilibrium phase transition.

\section*{Acknowledgements}
\NOTE{We thank three anonymous referees for their insightful and helpful comments, which have helped to improve this paper substantially.} This work was granted access to the HPC resources of MesoPSL financed by the R\'egion 
Ile de France and the project Equip@Meso (reference ANR-10-EQPX-29-01) of 
the programme Investissements d'Avenir supervised by the Agence Nationale pour 
la Recherche and the HPC resources of GENCI-TGCC-CURIE \& CINES	Occigen 
(Project No. A0050506421 \& A0070506421) where the present numerical simulations were performed. This work was also supported by the Agence nationale de la recherche (ANR DYSTURB project No. ANR-17-CE30-0004).

\section*{Declaration of interests}
The authors report no conflict of interest.

\appendix
\section{Heuristic derivation of the fast-rotating long box equations}\label{sec:appA}
In this appendix we present a heuristic derivation of the reduced equations discussed in the main text. A derivation based on the method of multiple scales is given in \citep{sprague2006numerical} for the Boussinesq equations, which reduces to our problem for vanishing density variations. \NOTE{The Navier-Stokes equation for a constant-density fluid in a reference frame rotating at the constant rate $\boldsymbol{\Omega} = \Omega \hat{e}_\parallel$ is}
\begin{eqnarray}
    \partial_t \mathbf{u} + \mathbf{u}\cdot \nabla \mathbf{u} + 2\Omega \hat{e}_\parallel\times\mathbf{u}=& - \nabla p + \nu \nabla^2\mathbf{u}+ \mathbf{f} \label{eq:RNS} \\
   \ \nabla \cdot \mathbf{u} =&0 \label{eq:inc},
\end{eqnarray}
where $\mathbf{u}=\mathbf{u}_\parallel + \mathbf{u}_\perp$ is velocity with $\mathbf{u}_\parallel = (\mathbf{u}\cdot \hat{e}_\parallel)\hat{e}_\parallel =u_\parallel \hat{e}_\parallel$ (we will use the same notation for all vectors), $p$ is pressure (divided by the constant density $\rho_0$) and $\mathbf{f}$ is the forcing. We impose triply periodic boundary conditions, the forcing is assumed to be solenoidal and to have zero average over the cuboid domain of dimensions $2\pi L\times2 \pi L \times2 \pi H$. We further restrict ourselves to a stochastic forcing injecting energy at a constant mean rate into both perpendicular and parallel motions $\langle \mathbf{f}_\perp \cdot \mathbf{u}_\perp \rangle =\langle f_\parallel u_\parallel  \rangle = \epsilon_{in}/2 \Rightarrow \langle \mathbf{f}\cdot\mathbf{u}\rangle  = \epsilon_{in}$, where $\langle \cdot \rangle$ denotes an ensemble average over inifinitely many realisations. The forcing is two-dimensional (independent of the parallel direction) and filtered in Fourier space to act only on a ring of perpendicular wavenumbers $\mathbf{k}$ centered on $|\mathbf{k}| = k_f=1/\ell_{in}$, precisely as considered in the main text.

\noindent Nondimensionalising (\ref{eq:RNS}, \ref{eq:inc}) using the perpendicular length scale $\ell_{in}$, the parallel length scale $H$ (in parallel derivatives), the timescale $(\epsilon_{in}^2/\ell_{in})^{1/3}$ and the velocity scale imposed by the forcing, $(\epsilon_{in} \ell_{in})^{1/3}$, gives 
\begin{eqnarray} 
    \NOTE{\partial_{\tilde{t}} \tilde{\mathbf{u}} + \tilde{\mathbf{u}}\cdot\tilde{\nabla} \tilde{\mathbf{u}} + \frac{2}{Ro} \hat{e}_\parallel \times \tilde{\mathbf{u}}} & \NOTE{= - Eu \tilde{\nabla} \tilde{p} + \frac{1}{\Rey} \tilde{\nabla}^2 \tilde{\mathbf{u}} + \tilde{\mathbf{f}}}, \label{eq:RNS_nd} \\
    \NOTE{\tilde{\mathbf{\nabla}}\cdot \tilde{\mathbf{u}}} & \NOTE{= 0,} \label{eq:inc_nd}
\end{eqnarray}
where $\tilde{\nabla} = \tilde{\nabla}_\perp + h^{-1} \tilde{\nabla}_\parallel$ and a tilde marks nondimensional quantities. In the above formula, $Ro = (\epsilon_{in} \ell_{in})^{1/3}/(\Omega \ell_{in})$ is the Rossby number, $Eu= P/(\epsilon_{in} \ell_{in})^{2/3}$ is the Euler number, $\Rey = (\epsilon_{in} \ell_{in})^{1/3} \ell_{in}/\nu$ is the Reynolds number and $h=H/\ell_{in}$ is the rescaled box height. Another nondimensional number is given by the rescaled box width $\Lambda=L/\ell_{in}$.  In the following, we shall omit tildes for simplicity.

\noindent 
Eliminating pressure from (\ref{eq:RNS_nd}) by applying the incompressible projection, defined for an arbitrary vector field $\mathbf{F}$ as $\mathbb{P}[\mathbf{F}]\equiv - \nabla^{-2} \nabla \times \nabla \times \mathbf{F} = \mathbf{F} - \nabla^{-2} \nabla (\nabla \cdot \mathbf{F})  $, $\nabla^2(\nabla^{-2} f) \nabla^{-2}(\nabla^2 f)= f$, and considering the equations for parallel velocity $u_\parallel$ gives
\begin{equation}
\partial_t u_\parallel  + \mathbf{u}\cdot\nabla u_\parallel  - h^{-1} \nabla^{-2} \partial_\parallel \lbrace \nabla\cdot(\mathbf{u}\cdot\nabla\mathbf {u})\rbrace + 2\lambda \nabla^{-2}\partial_\parallel\omega_\parallel = \frac{1}{\Rey} \nabla^2 u_\parallel + f_\parallel, \label{eq:vert_mom}
\end{equation}
and considering parallel vorticity $\omega_\parallel = \boldsymbol{\omega}\cdot \hat{e}_\parallel$, $\boldsymbol{\omega}=\nabla\times\mathbf{u}$ gives
\begin{equation}
\partial_t \omega_\parallel + \mathbf{u} \cdot \nabla \omega_\parallel \hspace{3.5cm}  - (2\lambda \hat{e}_\parallel+\boldsymbol{\omega})\cdot \nabla u_\parallel = \frac{1}{\Rey}  \nabla^2 \omega_\parallel + f_\omega, \label{eq:vert_vort}
\end{equation}
where $\partial_\parallel = \hat{e}_\parallel\cdot \nabla_\parallel$, $\lambda= (h Ro)^{-1}  = \ell_{in}^{5/3} \Omega /(\epsilon_{in}^{1/3} H)$ is identical to definition (\ref{eq:def_lamb}) in the main text and $f_\omega\equiv\hat{e}_\parallel\cdot (\nabla\times \mathbf{f})$. We consider the limit of simultaneously low Rossby numbers (fast rotation) and large aspect ratios, $h \equiv \epsilon^{-1} \gg 1$, $Ro = O(\epsilon) \ll 1$, such that $\lambda=O(1)$ (independent of $\epsilon$). This implies that $\nabla=\nabla_\perp +  \epsilon \nabla_\parallel $, such that $\nabla^2 = \nabla_\perp^2+O(\epsilon)$ and also $\nabla^{-2} = \nabla_\perp^{-2} + O(\epsilon)$. The fact that variations along the rotation axis are slow, meaning derivatives are $O(\epsilon)$, is a consequence of the Taylor-Proudman theorem, which is usually stated as forbidding fast variations in the limit $\Ro\to 0$, $h = O(1)$. Unlike in conventional quasi-geostrophy (in a thin layer) where parallel velocities are $O(\epsilon)$, both perpendicular and parallel velocities, as well as their perpendicular derivatives, are retained at leading order here. Nonetheless, just like in conventional quasi-geostrophy an important simplification arises from continuity,
\begin{equation}
\nabla\cdot \mathbf{u} = \nabla_\perp \cdot \mathbf{u}_\perp  + O(\epsilon)=0.
\end{equation} 
This means that the leading-order perpendicular velocity is incompressible and admits a streamfunction: $\mathbf{u}_\perp = \hat{e}_\parallel \times \nabla_\perp \psi$, hence $\omega_\parallel= \nabla_\perp^2 \psi$. Another simplification arises from the facts that
\begin{equation}
h^{-1}\nabla^{-2} \partial_\parallel \lbrace \nabla \cdot(\mathbf{u}\cdot\nabla\mathbf{u}))\rbrace =O(\epsilon)\ll \mathbf{u}\cdot \nabla u_\parallel =  \mathbf{u}_\perp\cdot \nabla_\perp u_\parallel + O(\epsilon).
\end{equation}
Finally, one finds that the vortex stretching term in the parallel vorticity equation (\ref{eq:vert_vort}) vanishes to leading order $\boldsymbol{\omega}\cdot\nabla u_\parallel = O(\epsilon)$. Combining these results yields the leading-order, asymptotically reduced governing equations 
\begin{eqnarray}
    \partial_t u_\parallel  + \mathbf{u}_\perp\cdot\nabla_\perp u_\parallel+ 2 \lambda \nabla^{-2}_\perp \partial_\parallel\omega_\parallel = \frac{1}{\Rey_\nu} \nabla^2_\perp u_\parallel+ f_\parallel,  \label{eq:red_vert_mom_app} \\
    \partial_t \omega_\parallel  +\mathbf{u}_\perp \cdot \nabla_\perp \omega_\parallel \hspace{0.675cm} - 2\lambda \partial\parallel u_\parallel = \frac{1}{\Rey_\nu} \nabla_\perp^2\omega_\parallel+f_\omega.\hspace{0cm} \label{eq:red_vert_vort_app}
\end{eqnarray}
Equations (\ref{eq:red_vert_mom_app}, \ref{eq:red_vert_vort_app}) are complemented by the geostrophic balance relation in the perpendicular direction, $\mathbf{u}_\perp = \hat{e}_\parallel \times \nabla_\perp \psi$, which implies $\omega_\parallel = \nabla_\perp^2 \psi$, making  (\ref{eq:red_vert_mom_app}, \ref{eq:red_vert_vort_app}) two equations for the two unknowns $u_\parallel$ and $\omega_\parallel$.
\bibliographystyle{jfm}
\bibliography{biblio}

\end{document}